\documentclass[aps,prx,showpacs,reprint]{revtex4-1}
\usepackage{amsmath}
\usepackage{amssymb}
\usepackage{setspace}
\usepackage{graphicx}
\usepackage{subfigure}
\usepackage{braket}
\usepackage{mathrsfs}
\usepackage[colorlinks = true,
            linkcolor = red,
            urlcolor  = blue,
            citecolor = blue,
            anchorcolor = blue]{hyperref}

\newcommand{\nn}{\nonumber \\}
\newcommand{\bs}{\boldsymbol}

\makeatletter

\newcommand{\Rom}[1]{\expandafter\@slowromancap\romannumeral #1@}
\makeatother

\begin{document}
\title{Ferromagnetism and glassiness on the surface of topological insulators}
\author{Chun-Xiao Liu}
\address{Condensed Matter Theory Center, Department of Physics, University of Maryland, College Park, Maryland 20742-4111, USA}

\author{Bitan Roy}
\address{Condensed Matter Theory Center, Department of Physics, University of Maryland, College Park, Maryland 20742-4111, USA}

\author{Jay D. Sau}
\address{Condensed Matter Theory Center, Department of Physics, University of Maryland, College Park, Maryland 20742-4111, USA}

\date{\today}

\begin{abstract}
We investigate the nature of the ordering among magnetic adatoms, randomly deposited on the surface of topological insulators. Restricting ourselves to dilute impurity and weak coupling (between itinerant fermion and magnetic impurities) limit, we show that for arbitrary amount of chemical doping away from the apex of the surface Dirac cone the magnetic impurities tend to arrange themselves in a spin-density-wave pattern, with the periodicity approximately $\pi/k_F$, where $k_F$ is the Fermi wave vector, when magnetic moment for impurity adatoms is isotropic. However, when magnetic moment possesses strong Ising or easy-axis anisotropy, pursuing both analytical and numerical approaches we show that the ground state is ferromagnetic for low to moderate chemical doping, despite the fragmentation of the system into multiple ferromagnetic islands. For high doping away from the Dirac point as well, the system appears to fragment into many ferromagnetic islands, but the magnetization in these islands is randomly distributed. Such magnetic ordering with net zero magnetization, is referred here as ferromagnetic spin glass, which is separated from the pure ferromagnet state by a first order phase transition. We generalize our analysis for cubic topological insulators (supporting three Dirac cones on a surface) and demonstrate that the nature of magnetic orderings and the transition between them remains qualitatively the same. We also discuss the possible relevance of our analysis to recent experiments.      
\end{abstract}

\maketitle

\section{Introduction}

Viewed from outside, a topologically nontrivial system encodes requisite (and possibly sufficient) information in the metallic surface/edge states to distinguish itself from trivial vacuum, occupying the external world. Existence of such gapless surface states is the hallmark signature of a topologically nontrivial phase of matter and cannot be eliminated unless the bulk of the system undergoes a topological phase transition. A celebrated example of such topologically nontrivial phase is the three dimensional strong $Z_2$ topological insulators (TIs) that supports odd number of massless Dirac cones on the surface~\cite{hasan2010colloquium, QiZhangRMP, fukane, rahulroy}. In nature such topologically nontrivial insulating phase can be found in strong spin-orbit coupled weakly correlated three dimensional semicondcutors~\cite{Hsieh:2008TI-ARPES, zhang2009topological, xia2009observation, chen2009experimental}, such as Bi$_2$Se$_3$, Bi$_2$Te$_3$, as well as in strongly correlated heavy fermion compounds~\cite{dzero2010topological, xu2013surface, neupane2013surface, jiang2013observation, alexandrov2013cubic, BRoySurfaceTKI}, such as SmB$_6$.

Since the successful discovery of three dimensional topological insulators in various strong spin-orbit coupled materials, manipulating the gapless surface by external magnetic field, ferromagnetic layer, magnetic doping has been an active field of research~\cite{axionfieldtheoryZhang, QAHERecentObservation,Yu61,Chang167,levchenko, carbotteTI, KerrMcdonald, ArmitageKerr, nagaosa, balatsky, abanin2011ordering, chen2010massive, hor2010development, liu2009magnetic, ye2010spin, garate2010magnetoelectric, EfimkinRKKY, PhysRevLett.109.107203}. Primary stimulation in this direction arises due to the possibility of observing, for example, quantum anomalous Hall effect~\cite{QAHERecentObservation,Yu61,Chang167,levchenko}, magneto electric effect~\cite{axionfieldtheoryZhang}, Faraday and Kerr rotation~\cite{axionfieldtheoryZhang, KerrMcdonald, ArmitageKerr}, which rely on the existence of fully gapped surface state (induced by a ferromagnetic order), achieved at the cost of breaking the time-reversal symmetry on the surface, while leaving the topologically nontrivial bulk band structure unharmed. Due to practical limitations, it seems most viable (experimentally) to stabilize a ferromagnetic order for itinerant surface states by injecting magnetic impurities on the surface, which has attracted ample attention in recent time~\cite{nagaosa, balatsky, abanin2011ordering, chen2010massive, hor2010development, liu2009magnetic, ye2010spin, garate2010magnetoelectric, EfimkinRKKY, PhysRevLett.109.107203}. A question of both fundamental and practical importance then arises naturally regarding the nature of the ordering among the magnetic impurities, when they are randomly deposited on the surface of a TI~\footnote{We here restrict ourselves to strong $Z_2$ TIs, supporting an odd number of surface Dirac cone. Nevertheless, our analysis can also be germane for the surface states of crystalline insulators, at least qualitatively.}. In this work we attempt to shed light on this issue by combining complimentary analytical and numerical analyses for the simplest realization of a three-dimensional TIs, supporting only one massless Dirac cone on the surface (germane to system like Bi$_2$Se$_3$) and cubic topological Kondo insulators (TKIs) (supporting three copies of massless Dirac cone on the surface). A schematic structure of the surface Brillouin zone for these two classes are shown in Fig.~\ref{fig_surfaceBZ}.

We here focus on dilute limit, when inter-impurity distance is larger than the lattice spacing so that we can safely neglect the direct interaction (Heisenberg type) between nearest-neighbor impurities. In this limit, the interaction among magnetic impurities is mediated by itinerant surface state, constituted by helical massless Dirac fermion, and is described by the Ruderman-Kittel-Kasuya-Yosida (RKKY) interaction~\cite{ruderman1954indirect, kasuya1956theory, yosida1957magnetic}. In general RKKY interaction is a rapidly oscillatory interaction at the scale of half of the Fermi wavelength ($\pi/k_F$). However, when the chemical potential is pinned at the apex of the surface Dirac cone (i.e. when $k_F = 0$), the RKKY interaction does not display any oscillation and the magnetic impurities are naturally arrange themselves in ferromagnetic pattern~\cite{liu2009magnetic}. Although such behavior of the RKKY interaction is singular, the resulting ferromagnetism is expected to stable against infinitesimal perturbation (such as change in chemical potential) for the following reason. When magnetic impurities arrange themselves in a ferromagnetic fashion, they in turn can produce a ferromagnetic order parameter for itinerant fermion, which then gaps out the Dirac point. Such effect has recently been demonstrated by a self consistent calculation~\cite{EfimkinRKKY}. Thus, unless the chemical potential is placed within the valence/conduction band, such ferromagnetic ordering should remain robust  and here we seek to understand the evolution of magnetic ordering among the impurities as the chemical potential is gradually tuned away from the Dirac point. However, weak fluctuations in the chemical potential on the scale of the gap caused by charge impurities are likely to destabilize this self-consistent effect which relies on the chemical potential being in the magnetic gap. In the following work we will assume that chemical potentials are sufficient to destroy strong selfconsistency effects. Our central results are the followings:

\begin{figure}
\begin{center}
\includegraphics[width=8cm, height=7cm]{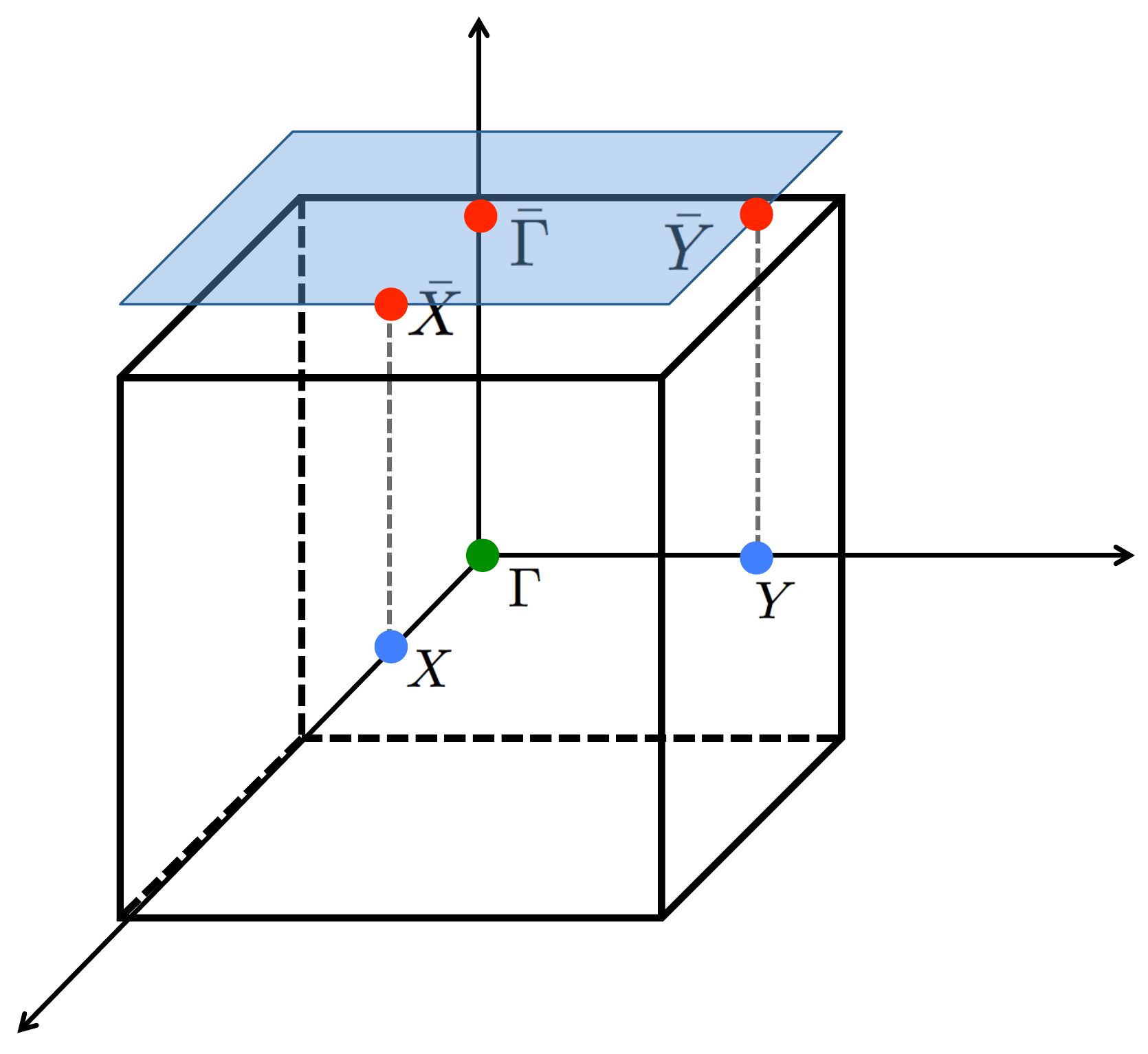} 
\caption{(Color Online) A schematic representation of surface Brillouin zone (blue shaded two dimensional object) and its connection to the high symmetry points in the bulk, where band inversion takes place. Often when bulk band inversion takes place at the $\Gamma$ point (green dot), such as in Bi$_2$Se$_3$, the surface Dirac cone is centered around $\bar{\Gamma}$ point. On the other hand, when bulk band inversion takes place at $X$, $Y$ and $Z$ points (blue dots), such as in SmB$_6$, three Dirac cones on (001) surface are located at $\bar{\Gamma}$, $\bar{X}$ and $\bar{Y}$ points. For brevity in the text of the paper, we will drop the \emph{bar} notation. }
 \label{fig_surfaceBZ}
\end{center} 
\end{figure}

\begin{enumerate}

\item{When chemical potential is tuned away from the surface Dirac point, the ground state of a collection of magnetic impurities sustains a spin-density-wave (SDW) pattern in weak coupling (among itinerant fermion and impurities) and dilute limit, with periodicity approximately $\pi/k_F$, if the magnetic moment of adatoms is \emph{isotropic}~\footnote{We here use the words pattern and ordering synonymously.}. }

\item{While such SDW pattern is quite generic on the surface of any TIs, the magnetic ordering on the surface of cubic TKIs display additional interesting features, when there exists a chemical potential imbalance between different Dirac cones~\footnote{This situation is quite generic since underlying cubic symmetry mandates that chemical potential at $X$ and $Y$ points is same, while displaying generic offset with the one at the $\Gamma$ point of the surface Brillouin zone.}. The SDW pattern on the surface of cubic TKIs displays two characteristic length scales or periodicities of oscillation, giving rise to \emph{beat}. The average chemical potential gives rise to periodicity of the overall modulation of SDW order, while the difference in the chemical potentials between Dirac cones located at $\Gamma$ and $X/Y$ points sets the periodicity inside each envelope of the SDW order (see Fig.~\ref{fig_surfaceBZ}). }

\item{ Typically the magnetic moment of higher spin impurity adatoms (such as Fe, Mn, Gd) possess strong Ising-like anisotropy. We show that such strong anisotropy in magnetic moment in turn gives rise to ferromagnetic ordering among magnetic impurities, at least when the chemical doping is not far away from the Dirac point. Through numerical analysis, we show that for small doping although the system breaks into multiple ferromagnetic islands. Ferromagnetic moment in each such island points in the same direction (although of different magnitudes) and system continues to sustain an overall net finite magnetization. This outcome is valid for the surface of TI as well as cubic TKI.}

\item{ By contrast, when chemical potential is tuned far away from the Dirac point, magnetization (an Ising variable) in these islands is  randomly distributed. The system then possesses net zero magnetization, giving rise to \emph{glassiness} on the surface of TIs or TKIs. More interestingly, the ferromagnetic and glassy phases are separated by a \emph{discontinuous} or \emph{first order phase transition}, which takes place when the characteristic length scale of the oscillation in the RKKY interaction is \emph{smaller} than the average inter-impurity distance. } 
      
\end{enumerate}

Let us now promote the organization principle for rest of the paper. In the next section (see Sec.~\ref{sec_ren}), we discuss the RKKY interaction among the magnetic impurities, mediated by surface Dirac fermions. In Sec.~\ref{sec_groundstates}, we analyze the arrangements among the magnetic impurities when the magnetic moment is isotropic as well as possesses strong Ising anisotropy. We present the numerical analysis, geared toward demonstrating the evolution of the magnetic order from low to high doping (away from the Dirac point) regime in Sec.~\ref{sec_dis}. We devote Sec.~\ref{sec_kondo} to generalize our analysis for the surface of cubic TKIs. Our findings are summarized in Sec.~\ref{sec_con}. Details of the ultraviolet regularization procedure in the derivation of RKKY interaction is presented Appendix~\ref{ren_app}.

\section{Spin susceptibility and RKKY interaction}~\label{sec_ren}

The spin susceptibility arising from itinerant fermions is capable of providing valuable insights into the nature of indirect exchange interaction among magnetic impurities, at least when they are placed far apart (dilute limit) and the interaction among them is only mediated by fermions. Therefore, by computing spin susceptibility one may also identify the nature of the magnetic ordering (such as paramagnetic or ferromagnetic) among doped magnetic impurities, with our focus here being on surface of TIs. Since we restrict ourselves to the dilute and weak coupling limit, the indirect exchange interaction can be extracted by employing the RKKY formalism~\cite{ruderman1954indirect, kasuya1956theory, yosida1957magnetic}.

The effective low-energy Hamiltonian, describing a helical metal on the surface of a three dimensional TIs is given by~\cite{hasan2010colloquium, QiZhangRMP}
\begin{align}
H_0 =  \sum_{\alpha \beta} \int d^2\bold{r} \; \Psi^{\dagger}_{\alpha}(\bold{r})  \left[ v_F( -i \hbar \nabla \times \bs{\sigma}_{\alpha \beta} ) \cdot \hat{z} - \mu \right] \Psi_{\beta}(\bold{r}),
\end{align}
where $\bs{\sigma}=(\sigma_x,\sigma_y)$ are standard Pauli matrices, $\Psi_{\alpha}(\bold{r})$ is the spinor wave function with spin projection $\alpha, \beta \in \{ \uparrow, \downarrow \}$ along the $z$ direction, $ v_F$ is the Fermi velocity of massless Dirac fermions and $\mu$ is the chemical potential, measured from the band touching point. The integral over $\bold{r}$ is restricted within the $xy$ plane, representing a surface of a three dimensional TI, and $\hat{z}$ points normal to such surface (see Fig.~\ref{fig_surfaceBZ}). Due to the underlying translational symmetry in the $xy$ plane the above Hamiltonian can also be represented as 
\begin{align}
H_0 = \sum_{\alpha \beta} \int^{\prime} \frac{d^2\bold{k} }{(2\pi)^2} \Psi^{\dagger}_{\alpha}(\bold{k}) \mathcal{H}^0_{\alpha \beta}(\bold{k})  \Psi_{\beta}(\bold{k}),
\end{align}
where the Hamiltonian operator reads as
\begin{align}
\mathcal{H}^0(\bold{k}) = \hbar v_F ( \bold{k} \times \bs{\sigma} )_z - \mu \label{bloch_ham},
\end{align}
where $\bold{k}=(k_x,k_y)$ and $k_j$s are spatial components of momentum. In what follows, we set $\hbar = 1$ and $v_F = 1$. Integral over momentum is restricted upto an ultraviolet cut-off $\Lambda_D$ (consult Appendix~\ref{ren_app} for details).

The spin susceptibility for such helical metal is defined as
\begin{align}
\chi^{ab}(\bold{r},\tau) & = - \braket{T_{\tau}\hat{S}^a(\bold{r},\tau)\hat{S}^b(\bold{0},0)}_0,
\end{align}
where $\braket{...}_0$ denotes the thermal average over the ensemble of free Dirac fermions and $a,b$ are the spin components. As a function of external frequency and momentum, the spin susceptibility becomes
\begin{align}
& \chi^{ab}(\bold{q},iq_n) = \int^{\beta}_0 d\tau \int d\bold{r} \; \chi^{ab}(\bold{r},\tau) \; e^{iq_n \tau - i \bold{q} \cdot \bold{r} } \nn
&= \sum_{m,n} \sum_{ik_n} \int d\bold{k} \; \mathcal{G}_{m,\bold{k}+ \bold{q}}(ik_n + iq_n) \; \mathcal{G}_{n,\bold{k}}(ik_n) \times \nn
& \braket{ u_{n , \bold{k}} | \sigma^a | u_{m , \bold{k} + \bold{q}} } \braket{   u_{m , \bold{k} + \bold{q}}     | \sigma^b |   u_{n , \bold{k}}  } 
\label{sus_basis_dept}
\end{align}
 where $n,m$ are band indices, $k_n, q_n$ are fermionic Matsubara frequencies, $ \beta = \frac{1}{k_B T}$ is the inverse temperature, and we here set $k_B=1$. The fermionic Green's function is $\mathcal{G}_{m,\bold{k}}(ik_n) = \left( ik_n - \epsilon_{m,\bold{k}} \right)^{-1}$. Now Eq.~(\ref{sus_basis_dept}) can be written more compactly as 
\begin{align}
&\chi^{ab}(\bold{q},iq_n) \nn
&=  \frac{1}{\beta} \sum_{i k_{n}} \int^{\prime} \frac{d^2\bold{k}}{(2\pi)^2} \mathbf{Tr} [\sigma^a \mathcal{G}(\bold{k} + \bold{q}, ik_n+iq_n ) \sigma^b \mathcal{G}(\bold{k} , ik_n ) ], \label{sus_q}
\end{align}
where $\mathbf{Tr}$ is operative over the spin idices and 
\begin{align}
\mathcal{G}(\bold{k} , ik_n ) & = \frac{1}{ik_n - \mathcal{H}^0(\bold{k})}  = \frac{ (ik_n+\mu) + ( \bold{k} \times \bs{\sigma} )_z }{(ik_n+\mu)^2 - \bold{k}^2}.\label{greens}
\end{align}
The integral over momentum is restricted by an ultraviolet cutoff $\Lambda_D$ up to which the dispersion of surface states is linear in momentum. We here focus only on the static part of the spin susceptibility, denoted as $\chi^{ab}(\bold{q}) \equiv \chi^{ab}(\bold{q},iq_n = 0)$.

\begin{figure}
\begin{center}
\includegraphics[width = \linewidth]{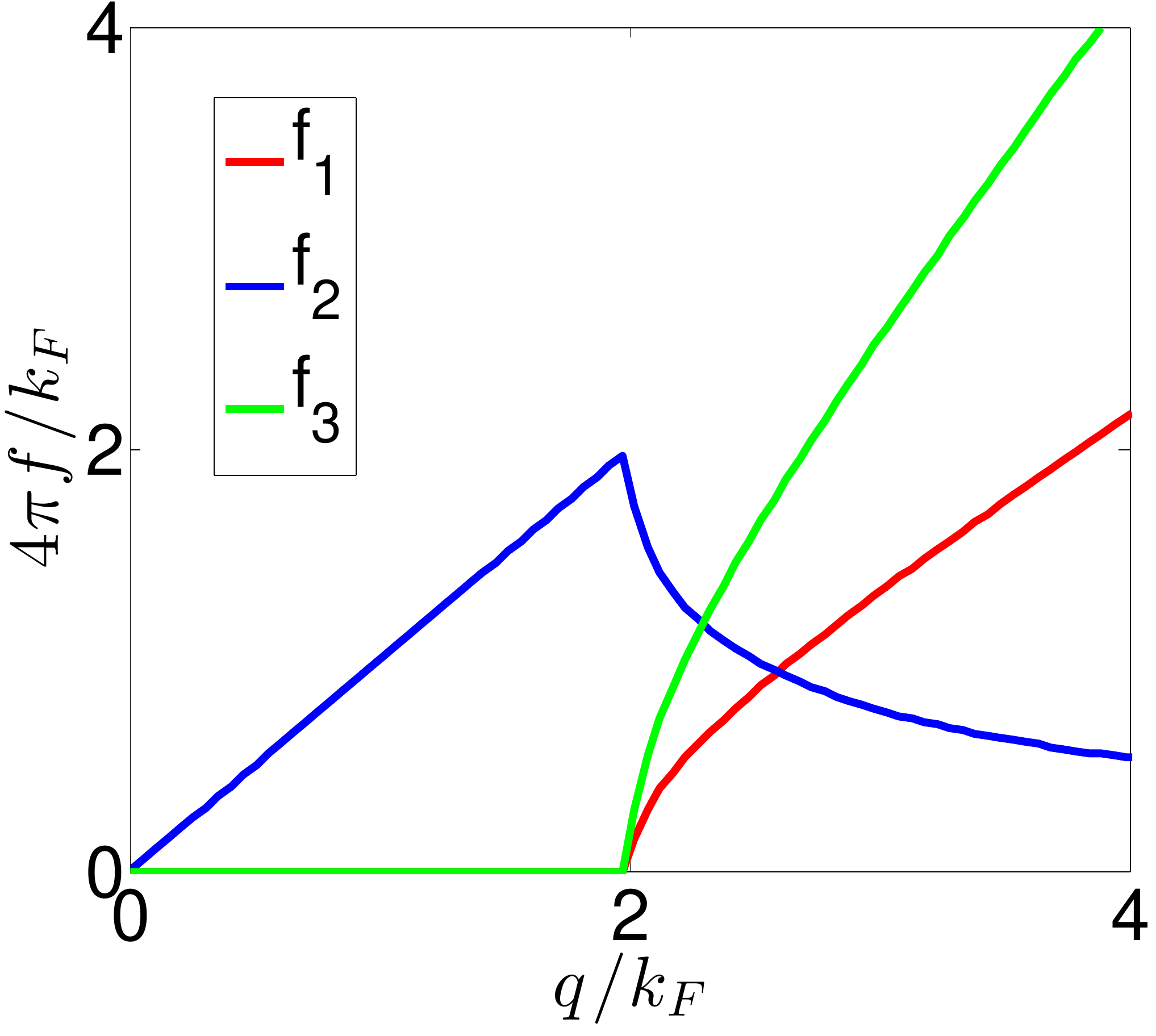} 
\caption{Scaling of $f_1,f_2$ and $f_3$ [in units of $k_F/(4 \pi)$], appearing in Eqs.~(\ref{ren_chi}) and (\ref{fs}), as a function of $q/k_F(=x)$. All of the three functions display \emph{discontinuity} at $q=2 k_F$, corresponding to the scale of Fermi wave vector.}
 \label{fig_f123}
\end{center} 
\end{figure}

As shown in Appendix~\ref{ren_app}, the diagonal components of $\chi^{ab}(\bold{q})$ display linear divergence with the ultraviolet cutoff $\Lambda_D$. Thus to remove such explicit cutoff dependence, we define a ultraviolet regularized spin susceptibility function according to 
\begin{equation}
\chi^{ab}_{ren}(\bold{q}) = \chi^{ab}(\bold{q}) - \chi^{ab}(\bold{0}). 
\end{equation}
A lengthy but straightforward calculation yields
\begin{align}
\chi^{ab}_{ren}(\bold{q})=
\begin{pmatrix}
f_1 \cos^2 \phi & \frac{f_1}{2} \sin 2 \phi & -i f_2 \cos \phi \\
\frac{f_1}{2} \sin 2 \phi  & f_1 \sin^2 \phi & -if_2 \sin \phi \\
  if_2 \cos \phi & if_2 \sin \phi & f_3  
\end{pmatrix} , \label{ren_chi}
\end{align}
where
\begin{align}
f_1 (x) &= \frac{|k_F|}{4 \pi} \mathrm{Re}\sqrt{1-x^2} + \frac{q}{8\pi} \mathrm{Re}\left[\sin^{-1} \sqrt{1-x^2}  \right] \nn
f_2 (x) & = \frac{q}{4\pi} \left( 1 - \mathrm{Re} \sqrt{1-x^2} \right), \nn
f_3 (x) &= \frac{q}{4 \pi} \mathrm{Re} \left( \sin^{-1} \sqrt{1 - x^2}\right),
\label{fs}
\end{align}
with $\bold{q} = q (\cos \phi , \sin \phi)$  and $x = 2k_F/q $. Explicit dependence of $f_j$s are shown in Fig.~\ref{fig_f123}. For brevity we dropped the explicit functional dependence of $f_j$s on $x=q/k_F$ from Eq.~(\ref{ren_chi}). The expression of these functions (namely $f_j$s) are different from the ones, announced previously in the literature~\cite{abanin2011ordering, garate2010magnetoelectric, EfimkinRKKY}. Such difference arises from appropriate ultraviolet regularization of leading order polarization bubble (see Appendix~\ref{ren_app}), which display \emph{linear} ultraviolet divergence due to the Dirac nature of underlying itinerant electrons.

To gain insight into the ground state configuration of magnetic impurities, we seek to find the effective Hamiltonian describing the exchange interaction among them. We here assume that helical Dirac fermion mediates indirect exchange coupling between two magnetic impurities. When magnetic impurities are deposited on the surface of a TI, one can treat each magnetic impurity as an external perturbation that couples to the spin degree of freedom of Dirac fermion through a point-like interaction  
\begin{align}
\hat{U}_{ext} = \lambda  \; \hat{ \bs{\sigma} }  \cdot \bs{S}(\bold{r}_i) \; \delta(\bold{r} - \bold{r}_i),
\end{align}
where $\lambda$ denotes strength of such interaction (dimensionless). We here assume that $\lambda \ll 1$ (placing the problem in the weak coupling regime), justifying a perturbative anaysis in powers of $\lambda$. In addition, we here treat impurity spin a classical quantity, which is a good approximation at least when the magnetic moment of dopant ions, such as the commonly used ones Fe, Mn, Gd, is large. The polarization of itinerant fermion at a given point $\bold{r}$ can then be quantified as 
\begin{align}
s^{a}_{\text{ind}}(\bold{r}) = \lambda \; \chi^{ab}(\bold{r} - \bold{R}_i ) \; S^{b}(\bold{R}_i),
\end{align}
where $s^{a}_{\text{ind}}(\bold{r})$ is the $a$-component of polarized spin of Dirac fermions, and $\chi^{ab}(\bold{r})$ is the spin susceptibility for Dirac fermion. Presence of another magnetic impurity at $\bold{R}_j$, interacting with Dirac fermion also causes polarization of itinerant spin at $\bold{R}_j$. Therefore, the exchange interaction between two magnetic impurities, located at $\bold{R}_i$ and $\bold{R}_j$ is given by (after integrating out massless Dirac fermion)
\begin{align}
H_{\text{eff}} = \lambda^2 S^{a}(\bold{R}_i) \chi^{ab}(\bold{R}_i - \bold{R}_j ) S^{b}(\bold{R}_j). \label{non_linear_ham}
\end{align}
Such indirect exchange interaction among local magnetic moments, mediated by itinerant fermions, is also known as RKKY interaction, with the non-linear constraint that the magnitude of each spin is fixed. It is worth mentioning that we here neglect classical and quantum fluctuations of spin since we are mainly interested in the ground state configuration of magnetic adatoms when they are deposited on the surface of TIs. We also neglect direct exchange interaction among magnetic ions, which can be a good approximation in the dilute limit. 

With the introduction of a constraint term the RKKY Hamiltonian is given by 
\begin{align}
 H_{RKKY} =  \lambda^2 \sum_{i \neq j}S^a(\bold{r}_i) \chi^{ab}(\bold{r}_i - \bold{r}_j )S^b(\bold{r}_j) \nn
+ \sum_i g ( [\bold{S}(\bold{r}_i)]^2 - 1 )^2 \label{impurity_ham}
\end{align}
where $a,b = x, y, z$ represents the three components of spin vector. The last term fixes the magnitude of each spin to be \emph{unity}, as g approaches infinity.

\section{variational analysis of a coarse grained model}~\label{sec_groundstates}

\begin{figure}
\begin{center}
\includegraphics[width = \linewidth]{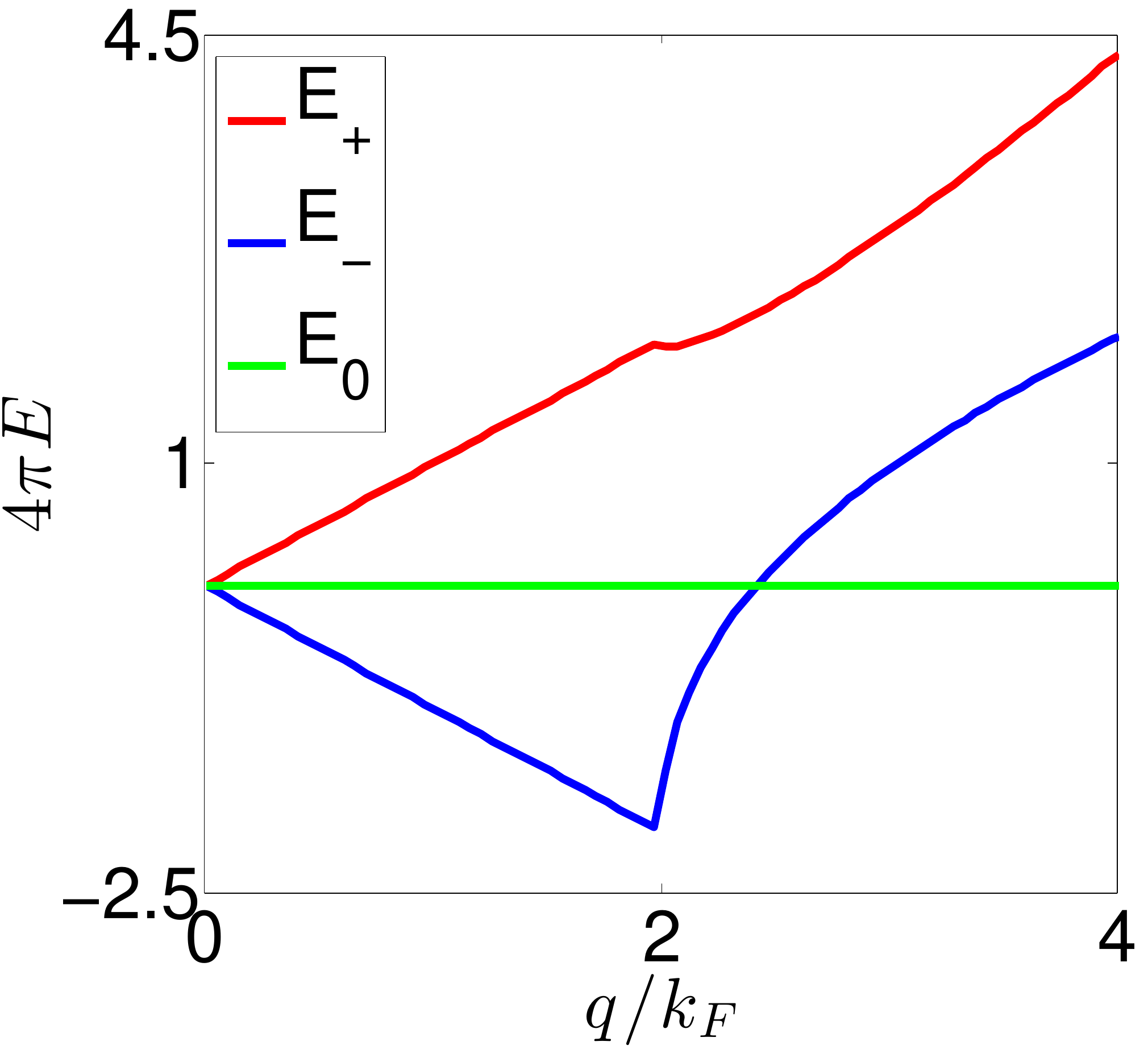}
\caption{Three branches of eigen energies for spin field obatined from the quadratic part of the Hamiltonian in Eq.~(\ref{ham_coarse}), assuming that magnetic moment does not possess any easy-axis anisotropy. Notably there is a global minimum at $q=2k_F$, which implies that in the ground state, the magnetic adatoms arrange themselves in a SDW pattern with wave vector $q = 2k_F$. }
\label{fig_eig_energy}
\end{center}
\end{figure}

In principle, one can search for the ground state configuration of magnetic impurities by minimizing the effective Hamiltonian, shown in Eq.~(\ref{non_linear_ham}). However, it is a challenging task due to the constraint of fixed magnitude, which leads to multiple local minima. Nonetheless, valuable insights into the actual ground state of the collection of magnetic impurities/spins can be achieved by pursuing a variational method and sacrificing the hard constraint over magnitude of the impurity spins, as we demonstrate below~\cite{cardy1996scaling} .
To soften the constraints on individual spins and also to reduce the effect of positional disorder of the spins, we define the spin field corresponding to magnetic impurities to be 
\begin{align}
\tilde{ \bold{S}}(\bold{r}) = \sum_i \bold{S}(\bold{r}_i) \delta(\bold{r} - \bold{r}_i ). 
\end{align} 
Within this representation, the exchange interaction term in Eq.~(\ref{non_linear_ham}) can be casted as
\begin{align}
& H_{S} = \lambda^2 \int d\bold{r}d\bold{r}' \tilde{S}^a(\bold{r}) \chi^{ab}(\bold{r} - \bold{r}' )\tilde{S}^b(\bold{r}') \nn
& = \lambda^2 \int^{\Lambda_{\chi}} \frac{d\bold{q}}{(2\pi)^2} \; \tilde{S}^{a}_{-\bold{q}} \;  \chi^{ab}(\bold{q})   \; \tilde{S}^b_{\bold{q}}.\label{effective_quadratic}
\end{align} 
The cutoff for the spin field in the momentum space ($\Lambda_\chi$) is assumed to be much smaller than that for massless Dirac fermion $(\Lambda_D)$, over which the dispersion is linear. The RKKY interaction kernel $\chi(\bold{r}-\bold{r}')$ favors a ferromagnetic alignment of spins at distances much shorter than the Fermi wave-length. Because of this, we can assume that the spin orientation varies slowly on the scale of the impurity spacing, which is assumed in this section to be much shorter than the fermi wave-length. Furthermore, the coefficient $\chi(\bold{r}-\bold{r}')$ in Eq.~(\ref{effective_quadratic}) can be assumed to be slowly varying in space on the scale of the impurity spacing for the same reason. Because of this, one may replace the spin field by a coarse grained spin field
\begin{align}
\bold{S}(\bold{r}) &= \sum_i \bold{S}(\bold{r}_i) e^{-\Lambda_{\chi}^2 (\bold{r} - \bold{r}_i)^2 } \nn
\bold{S}(\bold{q}) &= \tilde{\bold{S}}(\bold{q}) e^{-\frac{q^2}{\Lambda^2_{\chi}} }
\end{align}
As a result of such coarse-graining over the spin field, the stringent constraint over the magnitude of the spin field gets relaxed and the effective Hamiltonian in terms of the coarse-grained spin field is
\begin{align}
H_S = \lambda^2 \int^{\Lambda_{\chi}} \frac{d\bold{q}}{(2\pi)^2} S^{a}_{-\bold{q}}  \chi^{ab}(\bold{q})  S^b_{\bold{q}} +  g \int d\bold{r} [ ( S^z(\bold{r}))^2  - 1 ]^2, \label{ham_coarse}
\end{align}
where $g$ is now a \emph{finite} positive number.

\begin{figure}
\begin{center}
\includegraphics[scale=0.43]{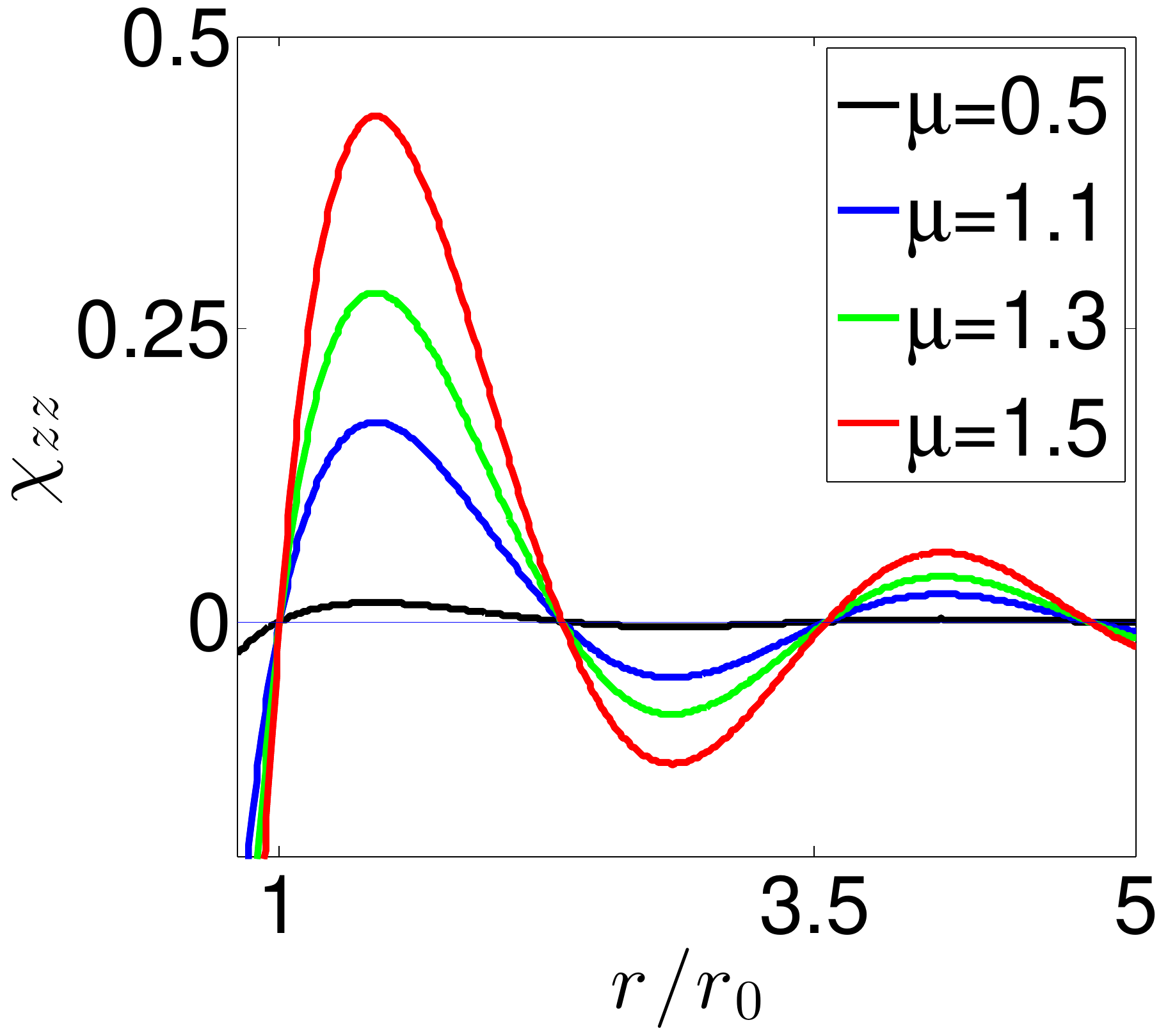}
\caption{The behavior of $\chi^{zz}(\bold{r})$ for different values of the chemical doping ($\mu$) in the zero temperature limit ($\beta \gg \mu$). The chemical potential $\mu$ is measured in units of $a^{-1}$, where $a$ is the average distance between adjacent magnetic impurities. Here, $r_0 \simeq 1.3 \mu^{-1}$, representing the length scale associated with the first zero of $\chi^{zz}(\bold{r})$. The ``wavelength" of the Bessel-like function $\chi^{zz}(\bold{r})$ is approximately $\lambda=2\pi/(1.3 r^{-1}_0) \simeq 2.5  r_0 \simeq \lambda_F / 2$, the characteristic length for the RKKY interaction. } 
\label{chi_zz_r}
\end{center}
\end{figure}

Before delving into the actual nature of the ground state configuration of magnetic impurities, we focus on the quadratic piece of the above Hamiltonian. Diagonalization of the quadratic Hamiltonian yields three energy eigenvalues, given by 
\begin{align}
E_{\pm} &= \frac{1}{2} \left[ (f_1 + f_3) \pm \sqrt{(f_1 - f_3)^2 + 4 f^2_2 } \right] \nn 
E_0 &= 0,
\end{align}
where $f_{1,2,3}$ are quoted in Eq.~(\ref{fs}). The momentum dependence of these three eigen energies are shown in Fig.~\ref{fig_eig_energy}, suggesting that there exists a global minimum at $q=2k_F$, indicating that at sufficiently low temperature the ground state of the collection of magnetic impurities is expected display a SDW order with wave vector $q_{SDW}= 2k_{F}$, if the magnetic moments are isotropic and thus can point in arbitrary direction.


\begin{figure*}[htbp]
\subfigure[]{
\includegraphics[width=5.5cm, height=5.25cm]{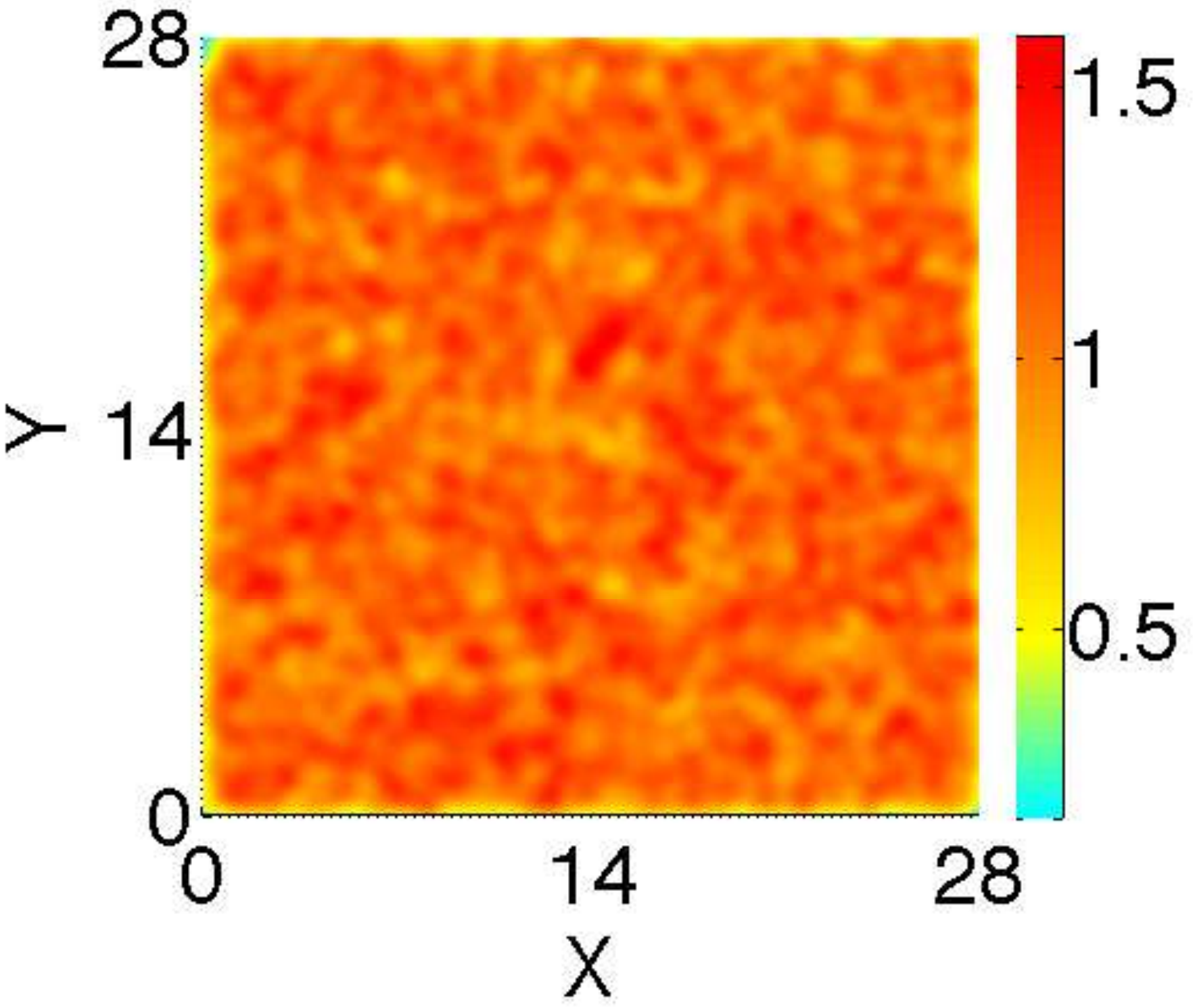}
\label{numerics_TI_a}
}
\subfigure[]{
\includegraphics[width=5.5cm, height=5.25cm]{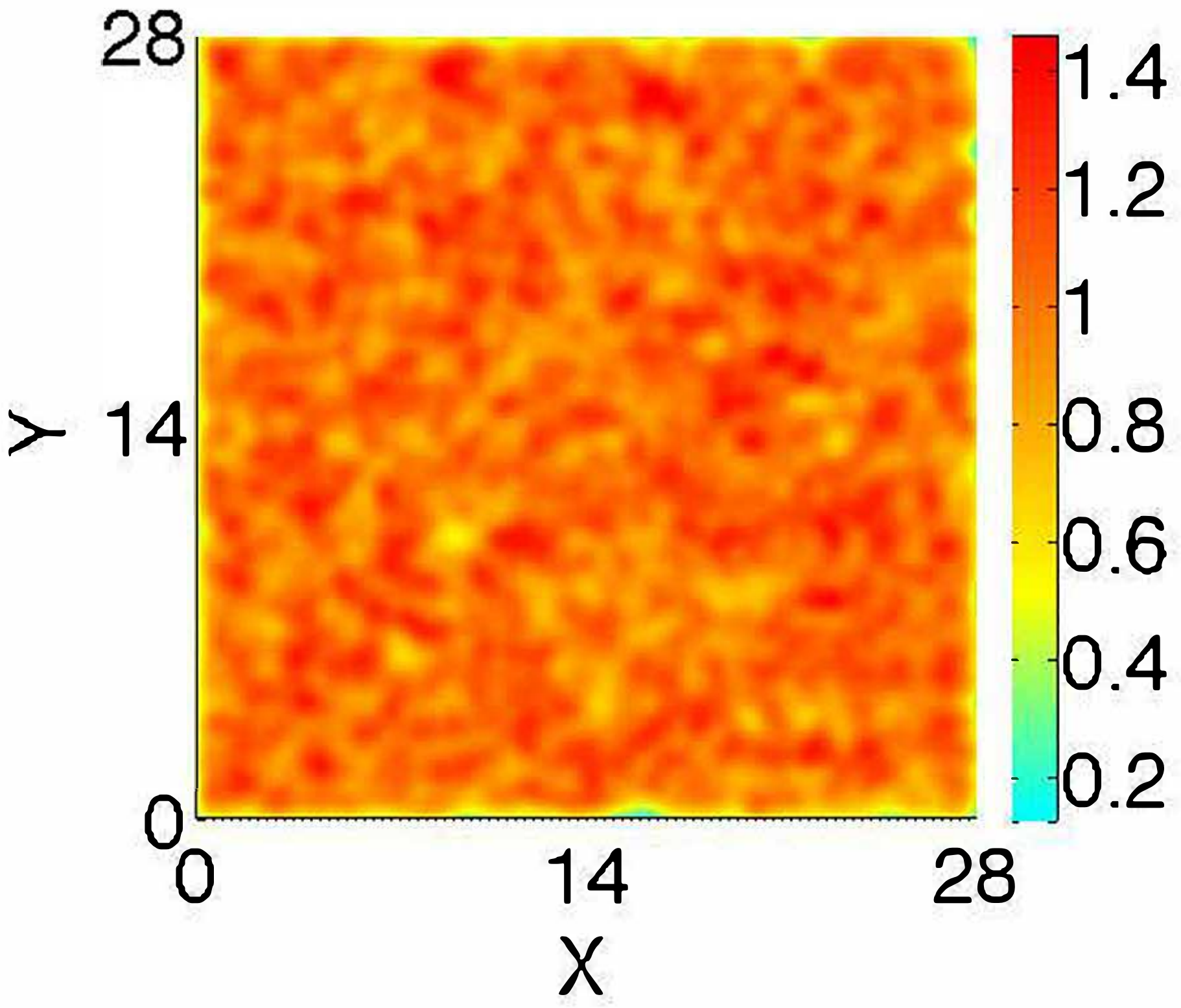}
\label{numerics_TI_b}
}
\subfigure[]{
\includegraphics[width=5.5cm, height=5.25cm]{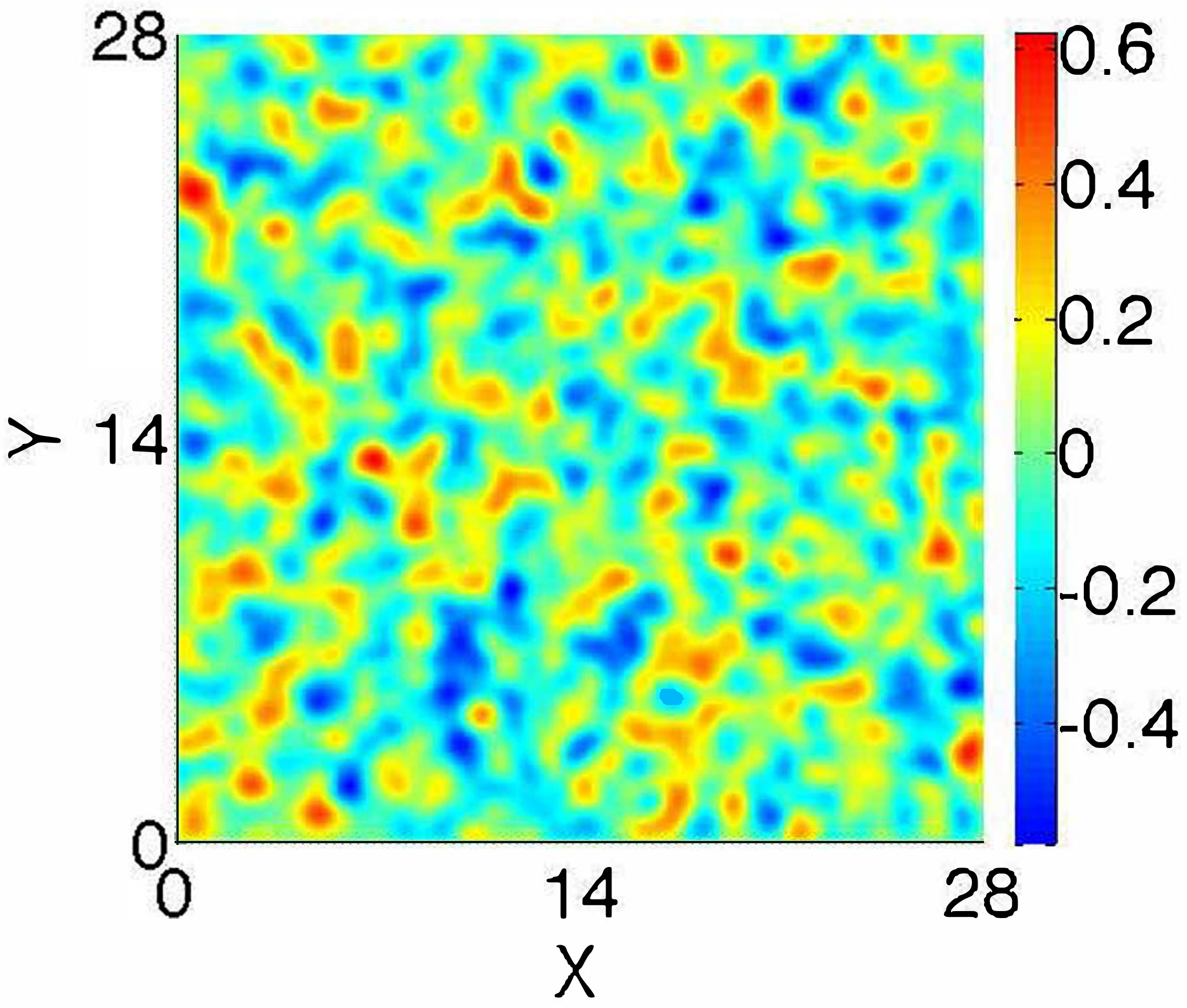}
\label{numerics_TI_c}
}
\caption{A disorder averaged (over 20 independent realization) plot for the ground state of impurity spin configuration on the surface of three dimensional topological insulators for (a) $\mu=0.5$ or $r_0=2.6$, (b) $\mu=1.1$ or $r_0=1.2$, (c) $\mu=1.5$ or $r_0=0.87$. Hence, for low [see (a)] and  moderate [see (b)] doping although the system break into multiple ferromagnetic islands, the magnetization in each such island points in the same direction, but they differ in magnitude. Consequently, system finds itself in ferromagnetic phase. On the other hand, for very high doping [see (c)] the magnetic moments in various islands are randomly oriented and system possesses net zero magnetization. In such a phase the system acquires glassiness. Even though the real configuration is spins located on discrete point-like positions, we smear them by a Gaussian function with width $W = 0.4$.}
\label{numerics_TI}
\end{figure*}


If, on the other hand, magnetic moments are Ising-like variables and point along the $z$-direction, there is only one branch of eigen energy with $E=f_3$. As shown in Fig.~\ref{fig_f123}, $f_3$ displays a plateau between $0 \leq q \leq 2k_F$ and the ground state configuration of magnetic impurities \emph{cannot be determined uniquely}. Hence, we need to account for the quartic term (soft constraint term after coarse-graining the spin field) to break such artificial degeneracy and pin the actual ground state. Thus with strong easy-axis anisotropy of the magnetic moment along the $z$-direction, we arrive at the phenomenological Landau free energy for the coarse-grained impurity spin field
\begin{align}
&F_{GL} =  \int d\bold{r} d\bold{r'} S^z(\bold{r}) \chi^{zz}_{ren}(\bold{r} - \bold{r'})S^z(\bold{r'}) \nn
& -m \int d\bold{r}  \left[ S^{z}(\bold{r}) \right]^2 + g \int d\bold{r}  \left[ S^{z}(\bold{r}) \right]^4  \nn
& =  \int_{\bold{q}}S^{z}_{-\bold{q}} [\chi^{zz}_{ren}(\bold{q}) - m ]S^z_{\bold{q}}  + g  \int_{\bold{p},\bold{q},\bold{k}} S^z_{\bold{p}} S^z_{\bold{q}} S^z_{\bold{k}} S^z_{-\bold{p}-\bold{q}-\bold{k}},
\label{Ising_free_energy}
\end{align}
which is one of the important results of this paper. Here $\chi^{zz}_{\text{ren}}$ is the $zz$-component of renormalized static spin susceptibility function (see Appendix~\ref{ren_app}). The ultraviolet cut-off dependence has been absorbed in the positive renormalized effective mass $m = 2g - \chi^{zz}_{\Lambda}(\bold{0})$, with$ \chi^{zz}_{\Lambda}(\bold{0})<0$. An unimportant constant has been dropped while arriving at the final expression in Eq.~(\ref{Ising_free_energy}). Next we compare the free energies with various trial ground states for magnetic impurities. Hence, the following analysis can be considered as variational approach to search for the best trial ground state.

Let us first consider a ferromagnetic order with 
\begin{align}
S_z(\bold{r}) = S_0.
\end{align} 
Plugging the above ansatz into Eq.~(\ref{Ising_free_energy}), we obtain the following free energy density
\begin{align}
f_{\text{FM}} = \frac{F_{\text{FM}}}{A} =  g S^4_0 + [\chi^{zz}_{ren}(0)-m]S^2_0,
\end{align}
where $A$ denotes the area of the two-dimensional surface of a TI. Notice that $\chi^{zz}_{ren}(0) = 0  < m $. Hence, the free energy with ferromagnetic background has lower free energy in comparison to that with an underlying disordered paramagnetic state, for which $S_0 = 0$ and the free energy is $f_{PM} = 0$. Minimizing the free energy with respect to the ferromagnetic order we obtain 
\begin{align}
S^2_0 = \frac{m-\chi^{zz}_{\text{ren}}(0)}{2g},
\end{align} 
and the corresponding free energy is given by 
\begin{align}
f^{\text{min}}_{\text{FM}} = -\frac{[\chi^{zz}_{\text{ren}}(0) - m]^2}{4g},\label{fm}
\end{align}
which is also a minima.

Next we consider a spin-density-wave ordering with unique wave vector $\bold{q} \neq 0$  
\begin{align}
S(\bold{r}) = S_0 \cos (\bold{q} \cdot \bold{r}).
\end{align}
Upon substituting the above ansatz into Eq.~(\ref{Ising_free_energy}), we find 
\begin{align}
f_{\text{SDW}} = \frac{F_{\text{SDW}}}{A} = \frac{3g}{8} S^4_0 + \left[ \frac{\chi_{\text{ren}}(q)}{2} - \frac{m}{2} \right]S^2_0.
\end{align}
For $\chi_{\text{ren}}(q)>2g > 0$, the paramagnetic phase with $S_0 = 0$ minimizes the free energy density (with $f_{\text{PM}}=0$). By contrast, for $ 0 \leq\chi(q) < 2g$, a SDW ordering with  
\begin{align}
S^2_0 = \frac{2[m-\chi_{\text{ren}}(q)]}{3g}, 
\end{align}
minimizes the free energy, and the minima of the free energy is given by
\begin{align}
f^{\text{min}}_{\text{SDW}} = -\frac{[\chi_{\text{ren}}(q) - m]^2}{6g}. \label{sdw}
\end{align}
Comparing Eq.(\ref{fm}) and Eq.(\ref{sdw}), we find that $f^{\text{min}}_{\text{FM}} < f^{\text{min}}_{\text{SDW}}$. Therefore, a ferromagnetic ordering is energetically superior over the paramagnetic as well as SDW states in the strong (Ising-like) anisotropic limit and low-doping regime.

\begin{figure}
\begin{center}
\includegraphics[scale=0.43]{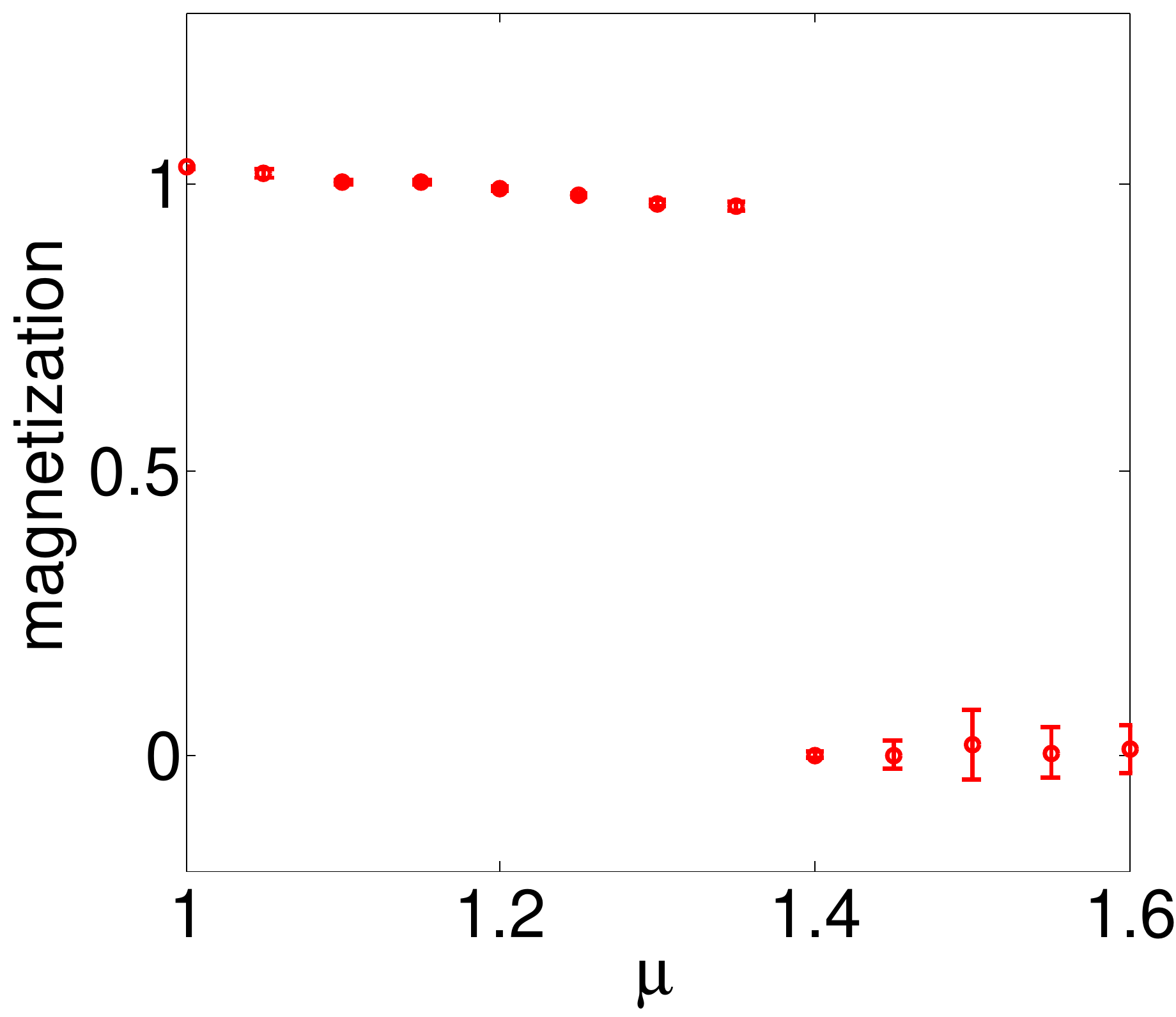}
\caption{ Disorder averaged net magnetization (normalized) as a function of chemical doping. Notice that across a critical chemical doping $\mu_{crit} \approx 1.3$ there is the first order phase transition between the pure ferromagnet and ferromagnetic spin glass phases. The normalized magnetization for low doping being slightly bigger than \emph{unity} is a consequence of softening the constraint due to the coarse grainign of the spin field. } 
\label{magnetization_TI}
\end{center}
\end{figure}

Finally, we consider a SDW ordering with multiple wave-vectors
\begin{align}
S(\bold{r}) = \sum^{N}_{n=1} S_n \cos(n \bold{q} \cdot \bold{r}),
\end{align}
for which the free energy density is given by 
\begin{align}
\tilde{f}_{SDW} = \frac{1}{2}\sum_n[\chi(nq)-m]S^2_n + \frac{g}{16}\sum_{m,n,l,p}S_nS_mS_lS_p \nn 
\times \sum_{i_n, i_m,i_l,i_p = 0,1} \delta[ (-)^{i_n} n +(-)^{i_m} m +(-)^{i_l} l +(-)^{i_p} p  ].
\end{align}
We then numerically search for the minimum of this free energy by using `fminunc' function in Matlab. For a specific choices of various parameters, namely $ g = 1, m = 2, \mu = 1, q=1, N = 6$, we search for the vector $(S_1, ..., S_6)^T$, yielding a minima of the free energy. We obtain $\tilde{f}^{min}_{SDW} = -0.8609$, while for same values of these parameters, $f^{min}_{FM} = -1, f^{min}_{SDW} = - 0.6667$. We also compared the free energy with various other choices of $q$, larger (smaller) than $k_F/2 (2 k_F)$. However, we always find $\tilde{f}^{min}_{SDW}>-1$. Thus, with strong easy-axis Ising anisotropic magnetic moment, the ferromagnetic order appears to be the most stable ground state. Next we examine the validity and robustness of ferromagnetic arrangement among the magnetic impurities in numerical simulation. 

\section{Numerical results  for single Dirac cone case}~\label{sec_dis}

The previous discussion on the nature of magnetic ordering on the surface of TIs based on the continuum theory is justified only in the low doping regime, where the Fermi wavelength ($\lambda_F$) is much longer than the average distance between adjacent magnetic impurities ($a$), i.e. $\lambda_F \gg a$. However, in high doping regime the notion of coarse grained spin breaks down and we need to numerically search for the magnetic ordering on the surface of TIs, as demonstrated below.

To carry out the numerical analysis, we first construct a system comprised of $800 (=N)$ Ising-like magnetic moments that are randomly distributed onto a two dimensional $R \times R$ square arena. Accordingly we choose $R=28$, so that the average distance between the nearest neighbor magnetic impurities is $a = \frac{R}{\sqrt{N}} \simeq 1$. Furthermore, we introduce a hard-core cutoff for the diatance between two impurities by setting $r_{min} \simeq 0.5$, ensuring that there is no clustering among magnetic impurities, in qualitative agreement with recent experiments~\cite{chen2009experimental}. Finally, we introduce a quartic term to constrain the magnitude of magnetic moments around same value, leading to the free energy for the system composed of a collection of magnetic impurities 
\begin{align}
F_{S} =  \sum_{i \neq j}S^z(\bold{r}_i) \chi^{zz}(\bold{r}_i - \bold{r}_j )S^z(\bold{r}_j) + g[ \left( S^{z}(\bold{r}_i) \right)^2-1]^2, \label{optimize}
\end{align}
where
\begin{align}
\chi^{zz}(\bold{r})  = -\frac{1}{\beta} \sum_{ik_n}\eta^2 \; \left[   K^2_0(\eta r) + K^2_1(\eta r)  \right],
\end{align}
with $\eta = \sqrt{(k_n-i\mu)^2}$, $k_n = \frac{(2n+1)\pi}{\beta}$~\cite{EfimkinRKKY}. The component of susceptibility along $z$ direction $\chi^{zz}(\bold{r})$ is assumed to be isotropic and its dependence on $r=|\bold{r}|$ is shown in Fig.~\ref{chi_zz_r} for various values of chemical doping $\mu$ [measured in units of $a^{-1}$, the inverse of average distance between adjacent magnetic impurities or the ultraviolet cut-off for the spin field, see Eq.(\ref{effective_quadratic})].

\begin{figure*}[htbp]
\subfigure[]{
\includegraphics[width=4.15cm, height=3.75cm]{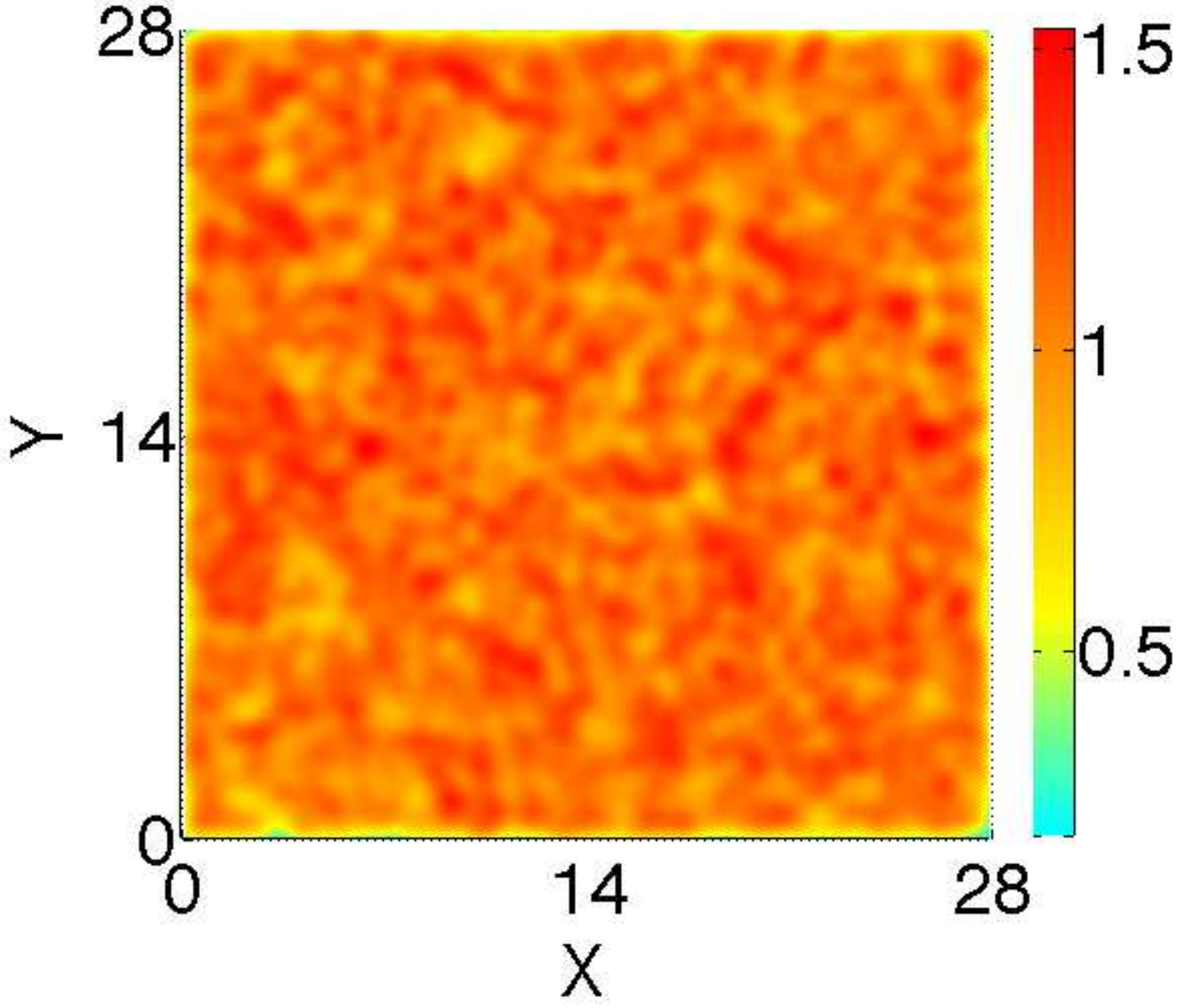}
\label{numerics_TKI_a}
}
\subfigure[]{
\includegraphics[width=4.15cm, height=3.75cm]{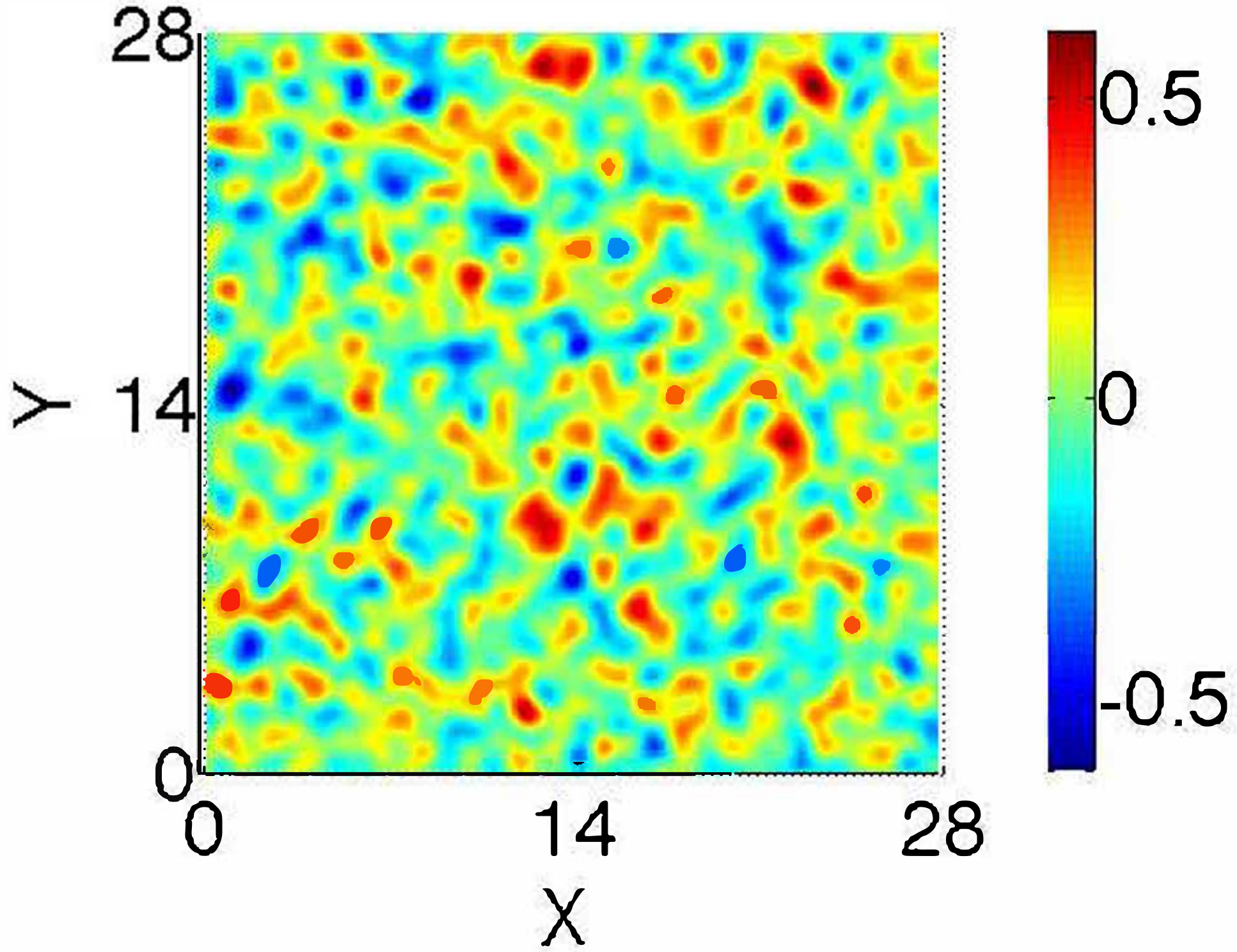}
\label{numerics_TKI_b}
}
\subfigure[]{
\includegraphics[width=4.15cm, height=3.75cm]{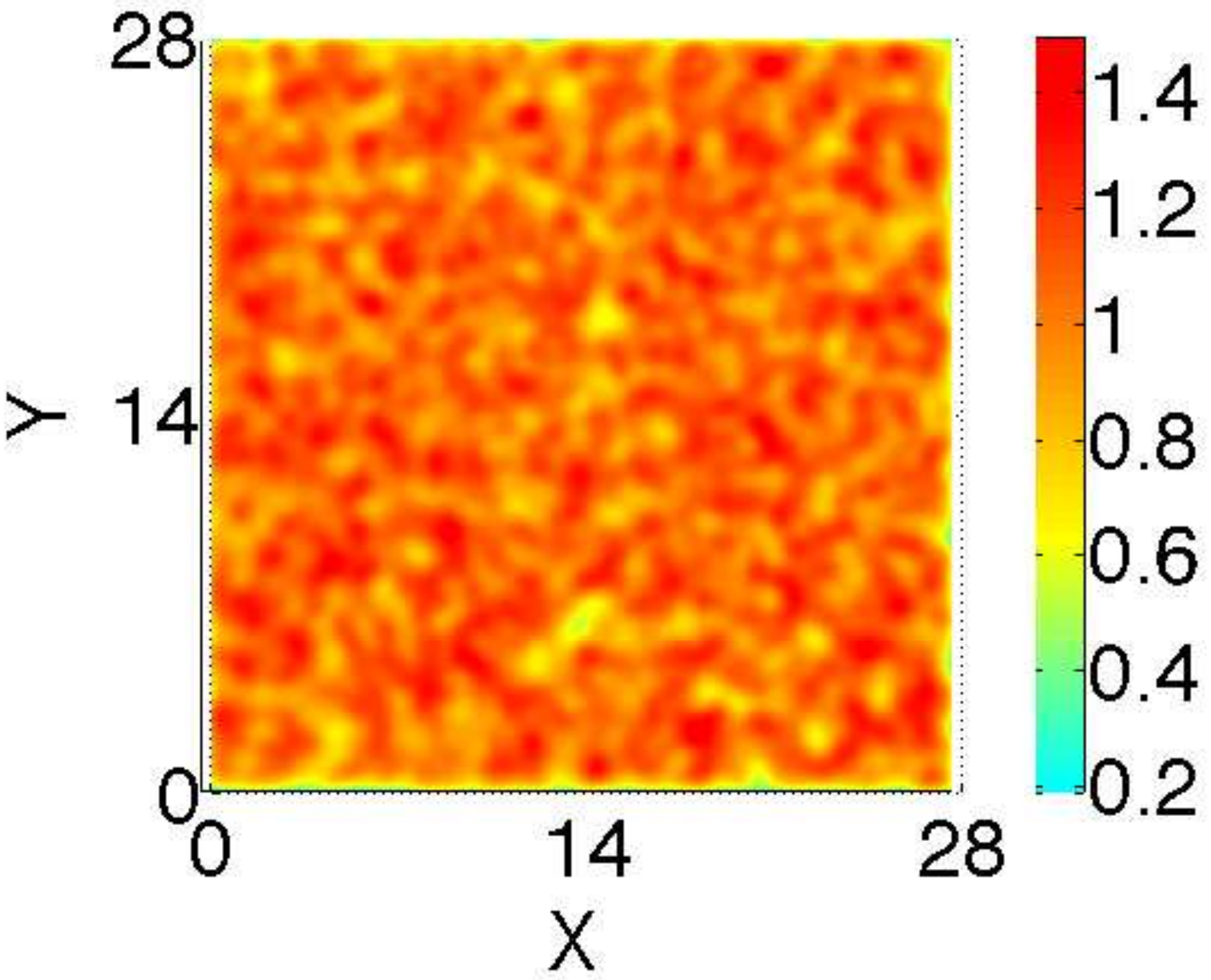}
\label{numerics_TKI_c}
}
\subfigure[]{
\includegraphics[width=4.15cm, height=3.75cm]{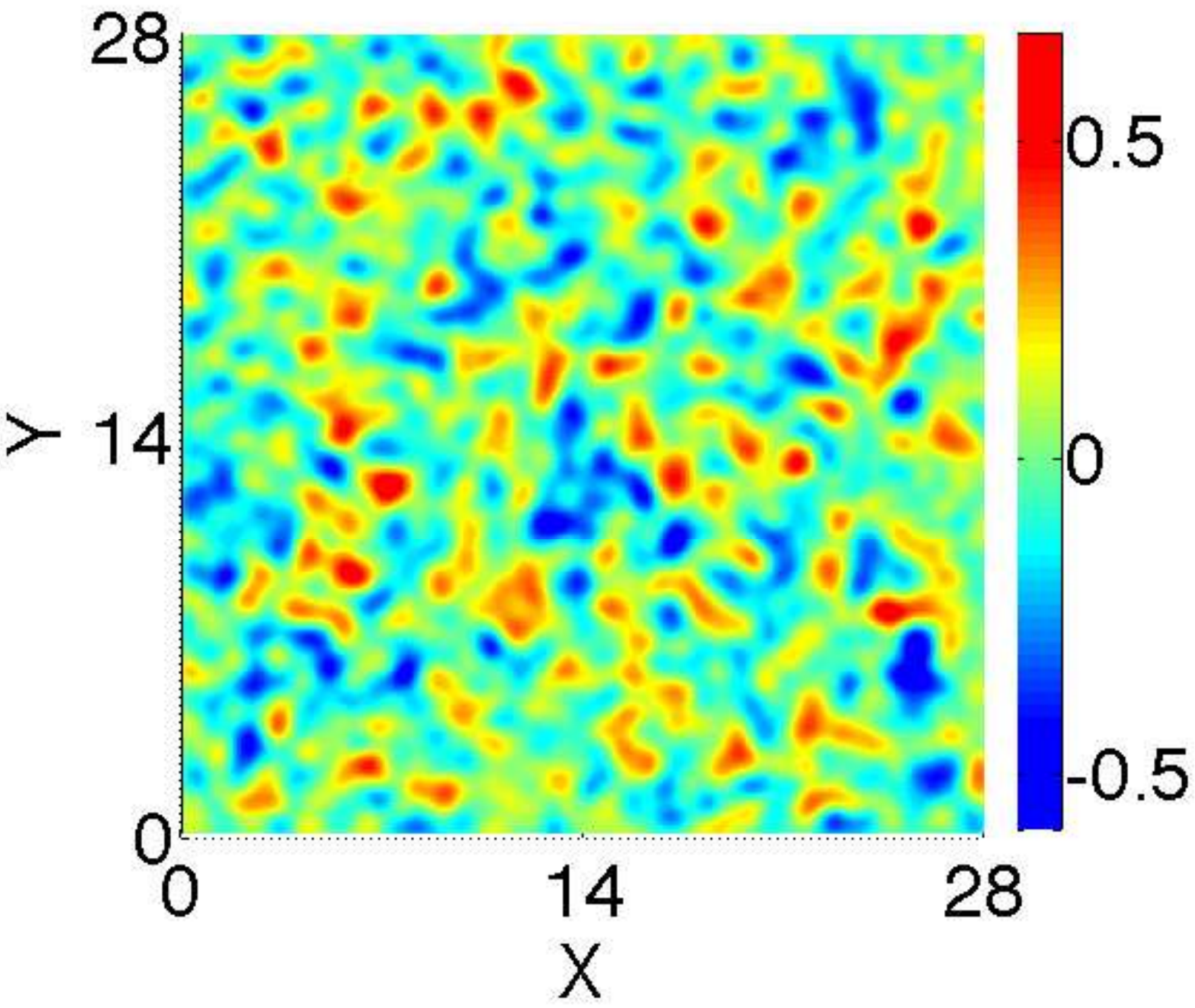}
\label{numerics_TKI_d}
}
\caption{ A disorder averaged (over 20 independent realization) plot for the ground state spin configuration for (a) $\mu_{\Gamma} = 0.5,\mu_X = \mu_Y = 1$, so that $r_{\text{eff}0} \simeq 1.56 > \braket{a}$ (low doping regime), (b) $\mu_{\Gamma} = 1.5,\mu_X = \mu_Y = 2$, so that $r_{\text{eff}0} = 0.71 < \braket{a}$ (high doping regime), (c) $\mu_{\Gamma} = 0.3, \mu_X = \mu_Y = 1.5$ with $r_{\text{eff}0} = 1.2 > \braket{a}$ (one in low doping regime while the other two are in high doping regime), (d) $\mu_{\Gamma} = 0.5, \mu_X = \mu_Y = 2$, but with $r_{\text{eff}0} = 0.87 < \braket{a}$ (still one in low doping regime while the other two are in high doping regime). The ground state for (a) and (c) are clearly ferromagnetic, while that in (b) and (d) displays glassiness.  }
\label{numerics_TKI}
\end{figure*}

Upon rescaling the distance $r$ by $r_0 \simeq 1.3 \mu^{-1}$, the zeros of $\chi^{zz}(\bold{r})$for all $\mu$ cross at particular points, where $r_0$ represents the value of $r$ where $\chi^{zz}(\bold{r})$ first undergoes a change in sign, see Fig.~\ref{chi_zz_r}. The ``wavelength" of the RKKY interaction is approximately given by $\lambda =2 \pi/(1.3 r^{-1}_0) \simeq 2.5  r_0 \simeq \lambda_F / 2$. As we will present in a moment that the relative strength of two length scales, namely $r_0$ and $a$, plays a crucial role in determining the actual nature of the magnetic ordering on the surface of TI. To demonstrate this competition we choose three particualr values of chemical doping $\mu = 0.5, 1.1, 1.5$, for which $r_0 \simeq 2.6 , 1.2 , 0.87$ respectively, allowing us the scan the magnetic ordering from low to high doping regime. We use the built-in function `fminunc' in Matlab to search for the minimum of the Free energy from Eq.~(\ref{optimize}). For all simulations we choose $g=5$, so that soft constraint condition is satisfied, i.e. $\delta |S| / \braket{|S|} \simeq 0.05$. The spin configuration, corresponding to the minima of the free energy, is shown in Fig.~\ref{numerics_TI}, for various values of $\mu$. Typically we average over 20 independent and random realizations of magnetic impurities.

Note that in the low doping regime (such as when $\mu=0.5$ for which $r_0 = 2.6 \gg \braket{a}$), the magnetic moments despite showing a spatial variation of average magnetic moment (still magnetization is $>0$ everywhere in the system), supports net finite magnetization, as shown in Fig.~\ref{numerics_TI_a}. Thus in the low doping regime, the magnetic ordering is \emph{ferromagnet}, in agreement with our previous analytical calculation. For moderately high doping (such as for $\mu=1.1$ for which $r_0 = 1.2 > \braket{a}$) the system breaks into several small islands, each of which supports net magnetization in the same direction, however of different magnitude, as shown in Fig.~\ref{numerics_TI_b} and the ground state is still ferromagnet. By contrast, for high enough doping (such as for $\mu=1.5$ for which $r_0 = 0.87 < \braket{a}$), the ground state configuration is composed of \emph{multiple} ferromagnetic islands. However, the relative orientation of magnetization in these islands are completely arbitrary and the system possesses net zero magnetization, as shown in Fig.~\ref{numerics_TI_b}. Such magnetic ordering qualitative mimics the structure of spin glass and we coin such phase as \emph{ferromagnetic spin glass}~\footnote{Here the word `glass' is used to describe a disordered phase, which is not a paramagnetic phase. But, such a phase does not break \emph{ergodicity}. }. Next we delve into the nature of the transition between the ferromagnet and ferromagnetic spin glass phases, across which the chemical potential ($\mu$) serves as a \emph{nonthermal} tuning parameter.

The nature of the magnetic phase transition, for example, can be pinned by studying the disorder averaged magnetization in the system. As shown in Fig.~\ref{magnetization_TI}, for low electron doping the surface of TIs possesses a net magnetization, which however smoothly decreases with increasing chemical doping. However, across a critical doping $\mu_{crit} \approx 1.3$ the magnetization drops abruptly and system enters into a phase where net magnetization is zero, the ferromagnetic spin glass (see Fig.~\ref{magnetization_TI}). Therefore, the zero temperature phase transition between these two phases is \emph{discontinuous or first order} in nature. Finally we come to the conclusion that when 
\begin{align}
\lambda_F = \frac{2 \pi}{1.3 r^{-1}_0} > 5 \braket{a},
\end{align}
 the ground state for anisotropic impurity spins is ferromagnet, while for $\lambda_F < 5 \braket{a}$ the ground state acquires glassiness and $\lambda_F \approx  5 \braket{a}$ represents the transition point between these two phases. Therefore, one can conclude that when the characteristic scale of oscillation for the RKKY interaction is bigger (smaller) than the average inter-impurity distance, the ground state is ferromagnet (ferromagnetic spin glass). Next we will generalize this observation for the surface states of cubic TKIs.

\section{Topological Kondo insulators}\label{sec_kondo}

So far we focused on the surface of topological insulators that supports only one two-component massless Dirac on surface. Such systems belong to class AII in ten fold way of classification. However, nontrivial AII invariant allows the existence of odd number of such flavor on the surface. Recent time has witnessed discovery of a TI that supports three copies of massless Dirac femrion, in the form of \emph{topological Kondo insulator} in SmB$_6$~\cite{dzero2010topological, xu2013surface, neupane2013surface, jiang2013observation, alexandrov2013cubic, BRoySurfaceTKI}. In the space group classification such TIs belongs to a distinct class $T-p3(4)_X$~\cite{juricic}. Recently there have been few experiments trying to explore the effects of depositing magnetic impurities of the surface of SmB$_6$~\cite{kim2013topological}. Here we explore possible magnetic ordering by accounting for an effective low energy model for the surface of cubic TKIs~\cite{BRoySurfaceTKI, SigristTKI, daiTKI, Vojta-2, BRoySurfaceInstability, BRoyHallTKI, Vojta-1}.   

To account for three Dirac cones located at the $\Gamma$, $X$ and $Y$ points of the surface Brillouin zone, we introduce the notion of valley indices and define a supervector as
\begin{align}
C_{\bold{k}} = (c_{\Gamma \bold{k} \uparrow},c_{\Gamma \bold{k} \downarrow},c_{X\bold{k}\uparrow},c_{X\bold{k}\downarrow},c_{Y\bold{k}\uparrow},c_{Y\bold{k}\downarrow})^T. 
\end{align}
The Hamiltonian in this basis reads as 
\begin{align}
H  =  \sum_{i=\Gamma,X,Y} \int_{\bold{k} \in \Omega} d\bold{k}H(\bold{k}-\bold{K}_i) 
=  \sum_{i} \int d\bold{k}H_i(\bold{k}),
\end{align}
where 
\begin{equation}
H_i(\bold{k}) = \bold{k} \times \vec{\sigma} - \mu_i,
\end{equation}
for $i=\Gamma, X$ and $Y$. Fermionic Green's function in this basis is block-diagonal and given by 
\begin{align}
\mathcal{G}(\bold{k}) = \mbox{diag.} 
\left( \mathcal{G}_{\Gamma}(\bold{k}), \mathcal{G}_X(\bold{k}), \mathcal{G}_Y(\bold{k}) \right),
\end{align}
whereas the general form of the spin operator is given by 
\begin{align}
\vec{\Sigma} = 
\begin{pmatrix}
\vec{\sigma} & b \vec{\sigma} & b \vec{\sigma}\\
 b \vec{\sigma} & \vec{\sigma} & a \vec{\sigma}\\
 b \vec{\sigma} & a \vec{\sigma} & \vec{\sigma}
\end{pmatrix}.
\end{align}
Two parameters $a$ and $b$ respectively denote the strength of \emph{inter-valley} scattering processes between $\Gamma$ and $X/Y$ points, and between $X$ and $Y$ points (see Fig.~\ref{fig_surfaceBZ}).

The total spin susceptibility for the surface states of cubic TKIs reads as 
\begin{align}
\chi^{ab}(\bold{q})  &= \frac{1}{\beta} \sum_{i k_{n}} \int \frac{d^2\bold{k}}{(2\pi)^2} \mathbf{Tr} \: \left[ \Sigma \mathcal{G}(\bold{k}+\bold{q}) \Sigma  \mathcal{G}(\bold{k}) \right] \nn 
&= \sum_i \chi^{ab}_{i}(\bold{q}) + \sum_{i \neq j} A_{ij} \chi^{ab}_{i,j}(\bold{q}),
\end{align}
where
\begin{align}
 \chi^{ab}_{i}(\bold{q}) &=\frac{1}{\beta} \sum_{i k_{n}} \int \frac{d^2\bold{k}}{(2\pi)^2} \mathrm{Tr} [\sigma^a \mathcal{G}_i(\bold{k} + \bold{q}, ik_n ) \sigma^b \mathcal{G}_i(\bold{k} , ik_n ) ], \nn
 \chi^{ab}_{i,j}(\bold{q}) &=\frac{1}{\beta} \sum_{i k_{n}} \int \frac{d^2\bold{k}}{(2\pi)^2} \mathrm{Tr} [\sigma^a \mathcal{G}_i(\bold{k} + \bold{q}, ik_n ) \sigma^b \mathcal{G}_j(\bold{k} , ik_n ) ],
\end{align}
and $A_{ij}  = b^2 \: \mbox{or} \: a^2$. Therefore, indirect exchange interaction (mediated by itinerant fermion) between two magnetic impurities is composed of two parts interaction mediated by (i) intra-valley scattering and (ii) inter-valley scattering (its strength is determined by coefficients $a$ and $b$). Thus understanding the nature of magnetic ordering is an interesting question, which can be of importance to recent and ongoing experiments on TKIs, such as SmB$_6$~\cite{kim2013topological}.

\begin{figure}
\begin{center}
\includegraphics[scale=0.43]{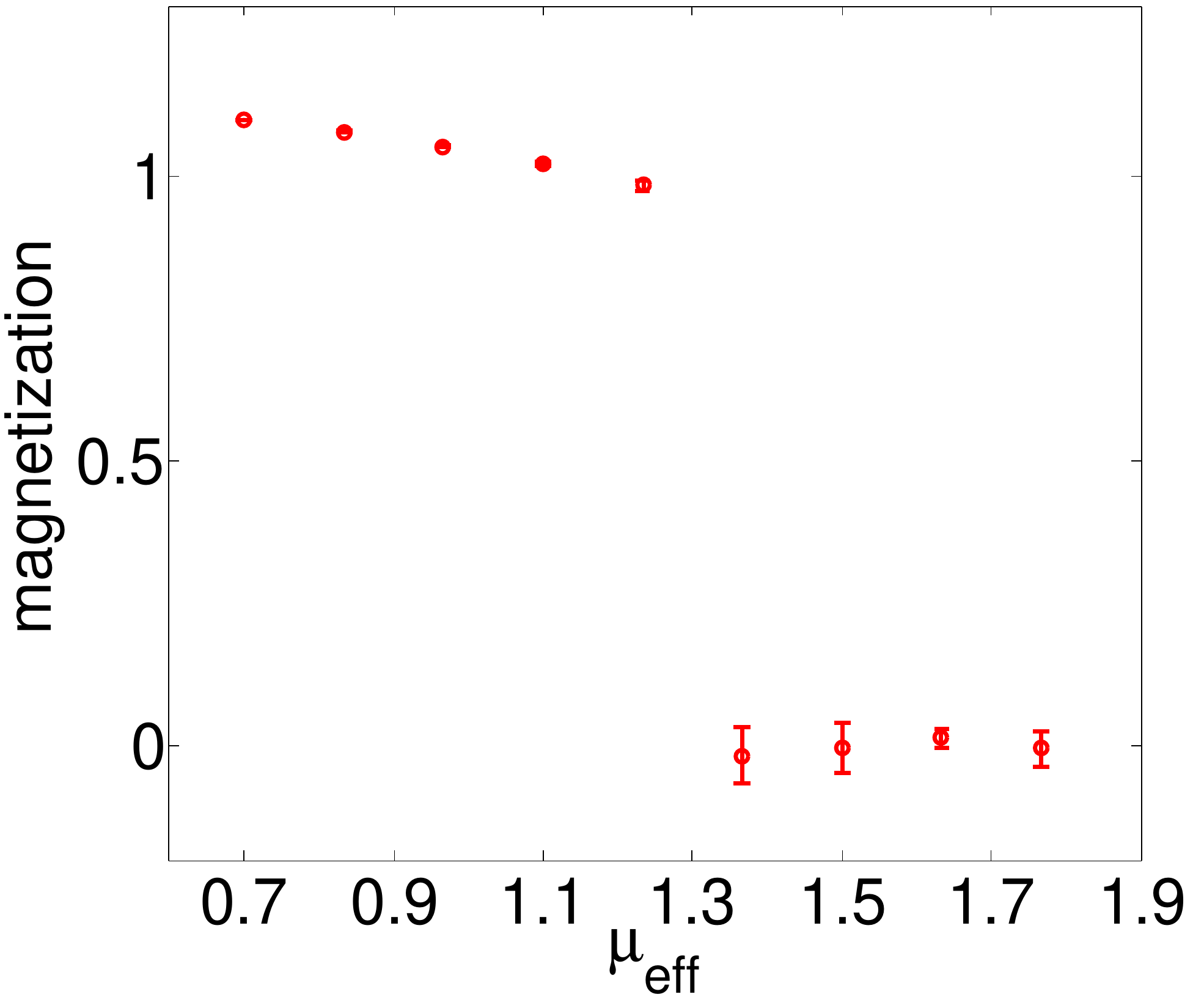}
\caption{ Disorder averaged net magnetization (normalized) as a function of effective chemical doping ($\mu_{eff}$) [see Eq.~(\ref{effective_chem})]. Notice that across a critical chemical doping $\mu_{crit} \approx 1.3$ a first order phase transition takes place between the ferromagnet phase and ferromagnetic spin glass state on the surface of cubic TKIs. The normalized magnetization for low doping being slightly bigger than \emph{unity} is a consequence of softening the constraint due to the caorse grainign of the spin field. } 
\label{magnetization_TKI}
\end{center}
\end{figure}

Let us first focus on a simpler situation by turning off the inter-valley scatterings (set $a=b=0$). Under this circumstance, the net spin susceptibility is a superposition for spin susceptibilities arising due to exchange interaction with fermions residing near $\Gamma$, $X$ and $Y$ valleys. Individually, the spin susceptibility functions have minima at wave vector $q=2 k^\Gamma_F, 2 k^X_F$ and $2 k^Y_F$, if the magnetic moment is isotropic (see Fig.~\ref{fig_eig_energy}). As a consequence, the magnetic impurities organizes in a spin density wave pattern that in addition displays \emph{beat}; with the larger wavelength (for the envelop) being inversely proportional to the difference of two Fermi wavevectors and the smaller wave length (determine the variation inside each such envelop) is set by the inverse of the algebraic mean of two Fermi wavevectors.   

When the magnetic moment bears strong Ising or easy-axis anisotropy along the $z$-direction, our previous discussion in the presence of a single Dirac cone can be generalized to gain insight into nature of magnetic ordering. Therefore, when the Fermi wavelengths of the three Dirac cones are all much larger than the inter-impurity distance, spin field can be coarse-grained and the ground state is expected to be ferromagnetic. On the other hand, when any of the three Fermi wavelengths is smaller than the inter-impurity distance, such analogy can no longer be established and we have to pursue numerical approach.

\begin{figure}[htbp]
\subfigure[]{
\includegraphics[width=4.0cm, height=4.0cm]{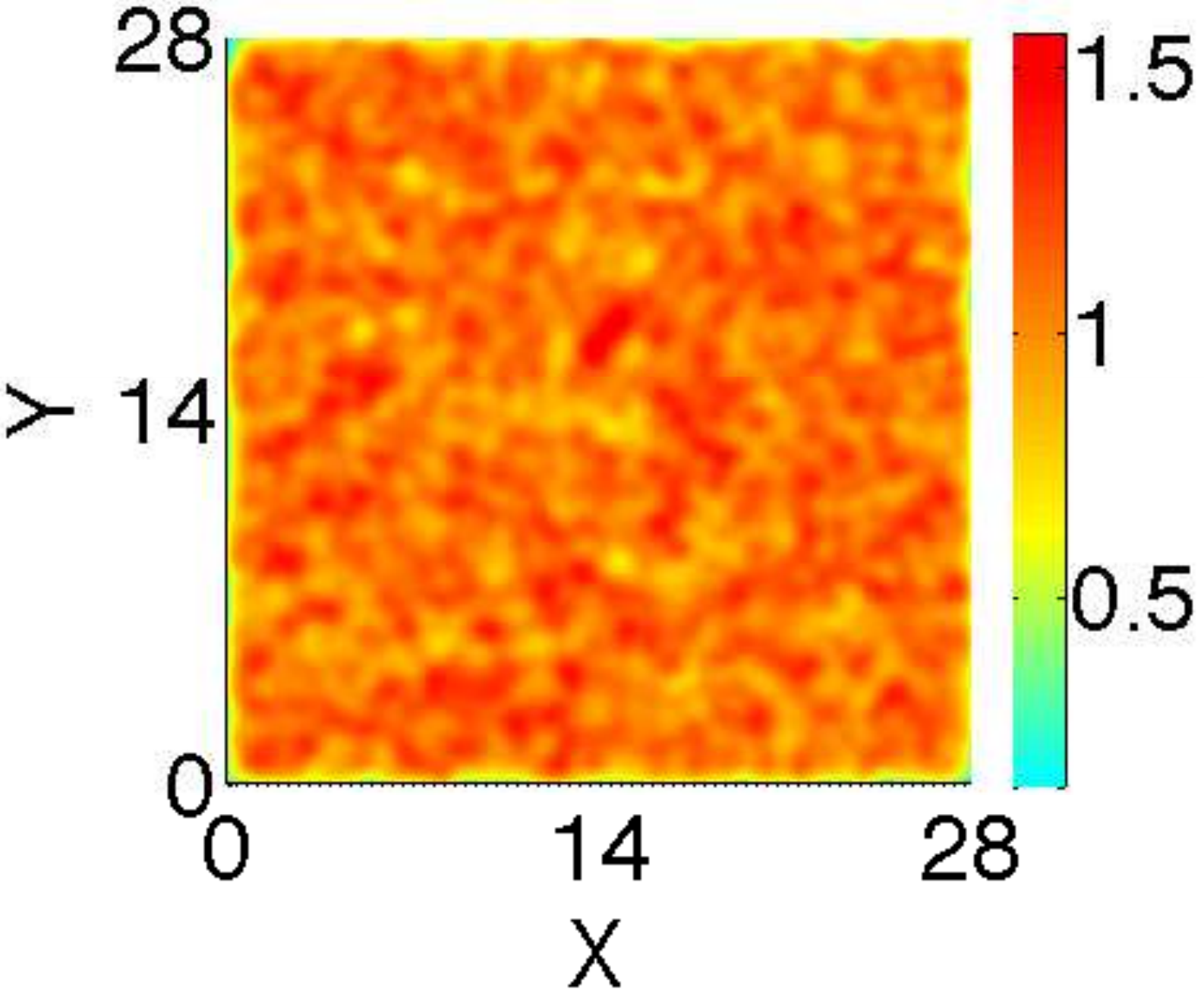}
\label{intervalley_a}
}
\subfigure[]{
\includegraphics[width=4.0cm, height=4.0cm]{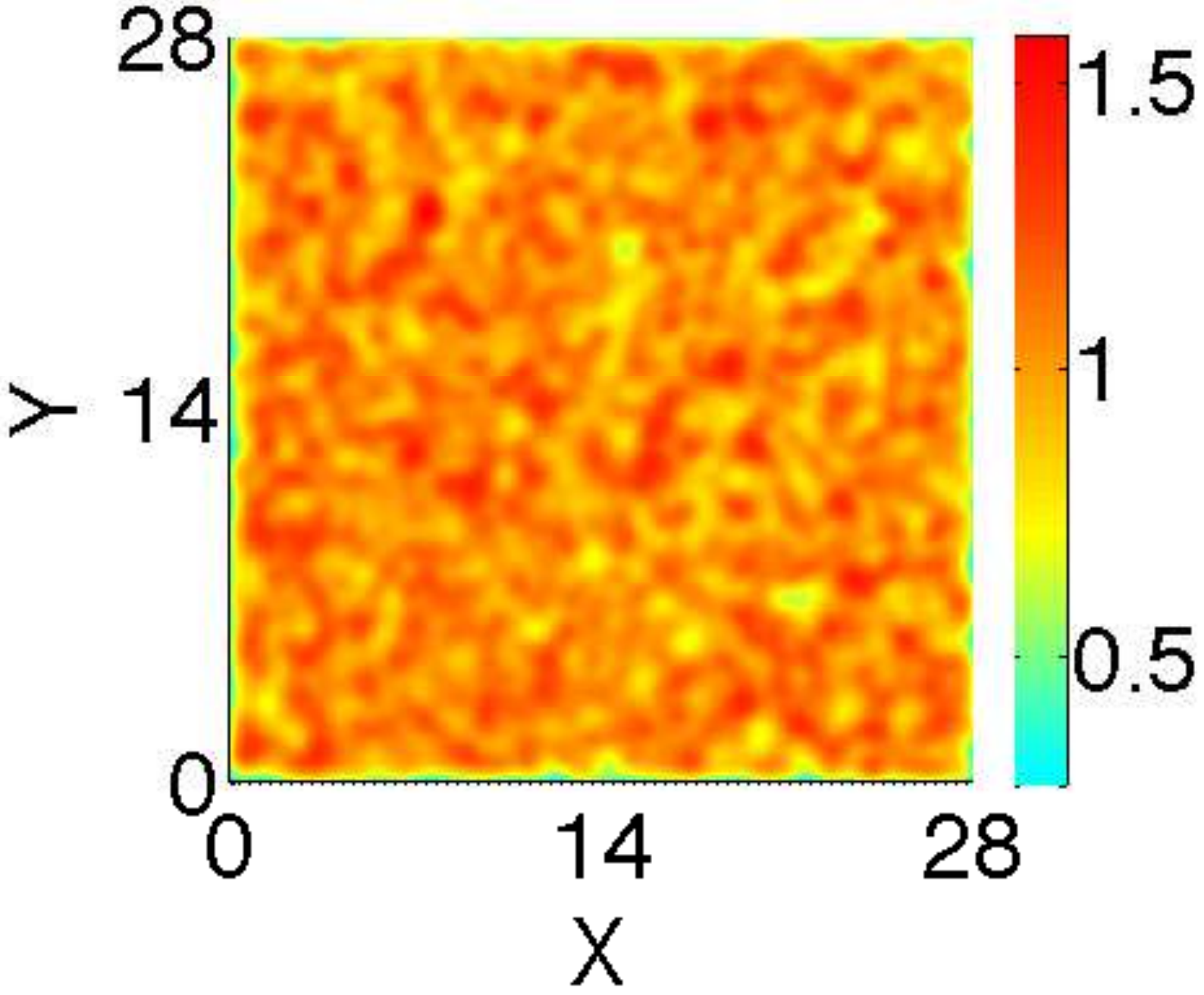}
\label{intervalley_b}
}
\caption{These figures are impurity spin configurations when inter-valley scattering are included, where we phenomenologically choose inter-valley scattering amplitude $a^2=1/3, b^2 = 1/5$. In Fig.~\ref{intervalley_a}, $\mu_{\Gamma} = 0.5, \mu_{X} =\mu_{Y} = 1$, while in Fig.~\ref{intervalley_b}, $\mu_{\Gamma} = 0.3, \mu_{X} =\mu_{Y} = 1.5$, which is the same as parameters set in Fig. \ref{numerics_TKI_a} and Fig. \ref{numerics_TKI_c}. We  can see that ferromagnetism is stable against moderate inter-valley scatterings. Each configuration is averaged over 20 independent disorder realizations.  }
\label{intervalley}
\end{figure}

The $zz$ component of the sipn susceptibility for cubic TKIs reads as
\begin{align}
&\chi^{zz}(\bold{r}) = \chi^{zz}_{\Gamma}(\bold{r}) + \chi^{zz}_{X}(\bold{r}) + \chi^{zz}_{Y}(\bold{r})
 + b^2 \Big( \chi^{zz}_{X\Gamma}(\bold{r})e^{-i \bold{K}_X  \cdot \bold{r}} \nn
&+ \chi^{zz}_{X\Gamma}(-\bold{r})e^{i \bold{K}_X  \cdot \bold{r}} 
 + \chi^{zz}_{Y\Gamma}(\bold{r})e^{-i \bold{K}_Y  \cdot \bold{r}} + \chi^{zz}_{Y\Gamma}(-\bold{r})e^{i \bold{K}_Y  \cdot \bold{r}}    \Big) \nn
& + a^2 \Big( \chi^{zz}_{XY}(\bold{r})e^{-i (\bold{K}_X -\bold{K}_Y )  \cdot \bold{r}}  + \chi^{zz}_{XY}(\bold{-r})e^{i (\bold{K}_X -\bold{K}_Y )  \cdot \bold{r}}    \Big) 
\end{align}
where
\begin{align}
\chi^{zz}_i(\bold{r})  &= -\frac{1}{\beta} \sum_{ik_n}\eta_i^2 [   K^2_0(\eta_i r) + K^2_1(\eta_i r)  ], \nn
\chi^{zz}_{ij}(\bold{r})  &= -\frac{1}{\beta} \sum_{ik_n}\eta_i \eta_j [   K_0(\eta_i r)K_0(\eta_j r) \nn
&+ K_1(\Lambda_i r)K_1(\Lambda_j r)  ], 
\end{align}
and $\eta_i = \sqrt{(k_n-i\mu_i)^2}$, $k_n = \frac{(2n+1)\pi}{\beta}$. Once again we generate 800 magnetic impurities on two dimensional $R \times R$ system, where $R=28$, so that the average distance between neighbor impurities $\langle a \rangle \simeq 1$. Notice that due to underlying cubic symmetry the chemical potential for the surface Dirac cones at $X$ and $Y$ points are same. For now we turn off all inter-valley scattering (by setting $a=b=0$). Results are displayed in Fig.~\ref{numerics_TKI}.

As we will demonstrate shortly that the nature of the magnetic ordering on the surface of TKIs can be anticipated by comparing an effective length scale for the RKKY interaction, given by $r_{\text{eff}0} \simeq 1.3 \mu^{-1}_{\text{eff}}$, where 
\begin{align}\label{effective_chem}
\mu_{\text{eff}} &= \frac{\mu_{\Gamma} + \mu_{X} + \mu_{Y}}{3}, 
\end{align}
with $\langle a \rangle$, the average distance between two nearest magnetic impurities. For example, when all three Dirac cones are in the low doping regime (with $\mu_{\Gamma} = 0.5 , \mu_{X,Y} = 1<\mu_{crit}=1.3$ and the corresponding $r_0 > \langle a \rangle$ individually), such that $r_{\text{eff}0}> \langle a \rangle$ our numerical simulation suggests that the ground state is ferromagnet, with net nonzero magnetization, as shown in Fig.~\ref{numerics_TKI_a}. By contrast, when all three Dirac cones are at high doping regime (with $\mu_{\Gamma} = 1.5 , \mu_{X,Y} = 2$), so that $r_{\text{eff}0} = 0.71 < \braket{a}$, the spin configuration in the ground state fragments into mutiple islands, with random orientation of magnetization, such that system possesses net zero magnetization, representing the ferromagnetic spin-glass-like phase, as shown in Fig.~\ref{numerics_TKI_b}. These two situations can be considered as generalization of the situation with single Dirac cone. However, a more interesting situation arises when the doping concentration for different Dirac cones are different. Such a situation is conceivable and can also be realized in experiments due to the generic offset among the energy of the Dirac points located at $\Gamma$ and $X/Y$ points~\cite{BRoySurfaceTKI, BRoySurfaceInstability, BRoyHallTKI}. The underlying cubic symmetry pins the Dirac cone at $X$ and $Y$ points at the same energy, which are generically different from the one at the $\Gamma$ point. Let us consider a situation when $\mu_{\Gamma} = 0.3 , \mu_{X,Y} = 1.5$, i.e. Dirac cone at $\Gamma$ point is at low electron-doping regime, while those at $X,Y$ points are at high electron-doping regime. With such choices of the parameters $r_{\text{eff}0} = 1.2 > \braket{a}$ and our numerical analysis suggests that the ground state in ferromagnet, see Fig.~\ref{numerics_TKI_c}. Finally, we set $\mu_{\Gamma} = 0.5 , \mu_{X,Y} = 2$, i.e. Dirac cone at $\Gamma$ point is at low electron-doping regime, while those at $X,Y$ points are at high electron-doping regime, for which $r_{\text{eff}0} = 0.87 < \braket{a}$. Numerical analysis suggest that the ground state with these choices of the parameter is ferromagnetic spin glass, as shown in Fig.~\ref{numerics_TKI_d}. Thus, our numerical analysis strongly suggests that when the effective zero point for $\chi^{zz}(r)$, namely $r_{\text{eff}0}$, is greater (smaller) than the average nearest neighbor distance, the ground state for impurity spins is ferromagnet (ferromagnetic spin glass).

By computing the disorder averaged net magnetization in the system, we can track the nature of the transition between a ferromagnet and the ferromagentic spin glass phases. As shown in Fig.~\ref{magnetization_TKI}, for small $\mu_{eff}$ the system is ferromagnet, which at larger $\mu_{eff}$ system displays glassiness. Around a critical strength of effective chemical potential defined in Eq.~(\ref{effective_chem}), namely $\mu_{eff} \simeq 1.3$ the system undergoes a first order phase transition. 

Finally, we take into account inter-valley scattering and in particular seek to investigate the stability of ferromagnetic arrangement of magnetic impurities against the onslaught of inter-valley scattering. We choose the following parametrization for inter-valley scattering $a^2 = \frac{1}{3}$ (representing the strength of scattering between $\Gamma$ and $X/Y$ valleys) and $b^2 = \frac{1}{5}$ (capturing the strength of scattering between $X$ and $Y$ valleys), while other parameters are kept same as those in Fig.~\ref{numerics_TKI_a} and Fig.~\ref{numerics_TKI_c}. The relative strength of $a$ and $b$ is roughly proportional to the ratio of the separation between $\Gamma$ and $X/Y$ valleys, and $X$ and $Y$ valleys. For these choices of parameters, the spin configuration in the ground state is displayed in Fig.~\ref{intervalley}, and we find that the ferromagnetic arrangement among the magnetic impurities can be robust against the inter-valley scattering.

It should be noted that we here completely neglect the effects of residual electron-electron interaction on the surface of cubic TKIs. Since, the bulk band inversion in these systems takes place through the hybridization among $d$ and $f$ electrons, the surface state is also composed of linear superposition of these two orbital and can constitute a strongly correlated Dirac liquid. Strong interaction among the surface states can lead to various exotic phases among which spin liquid~\cite{nokolic, sachdev}, broken symmetry phases~\cite{BRoySurfaceInstability, BRoyHallTKI}, chiral liquid~\cite{onur} have been proposed theoretically. However, at this stage it is not clear how strong is the residual electronic interaction on the surface. At least, for sufficiently weak interaction our proposed phases (pure ferromagnet and ferromagentic spin glass) should be robust. Nevertheless, effects of electronic interaction should now be systematically incorporated to test the regime of validity of our analysis (see Ref.~\cite{allerdt2016competition} for similar discussion relevant to magnetically doped graphene), which, however, goes beyond the scope of present discussion.

\section{Summary and discussion}\label{sec_con}

To summarize, pursuing complementary analytical and numerical analyses, we here investigate the nature of magnetic ordering on the surface of simple topological insulators (containing only one flavor of two component Dirac fermion) and cubic topological Kondo insulators (supporting three copies of two component Dirac fermion), when magnetic impurities are randomly deposited. We here work in the dilute magnetic impurity limit so that direct exchange interaction can be neglected and interaction among two impurities is mediated by surface itinerant fermions (but the coupling between these two degrees of freedom is small). Such indirect interaction among magnetic impurities assumes the form of a RKKY interaction. We show that when magnetic moment of impurity adatom is isotropic and the chemical potential is pinned away from the Dirac point, the ground the on the surface of conventional topological insulators is a spin-density-wave with wavelength approximately $\pi/k_F$. On the other hand, due to a generic offset among the energy of three Dirac points on the surface of cubic topological insulators, a similar spin-density-wave arrangement assumes the profile of a \emph{beat}, with two distinct wavelengths determining the short and large length scale behaviors.

The situation gets quite involved when magnetic moment bears strong Ising-like or easy-axis anisotropy. For low chemical doping, performing coarse graining over the impurity spin field, we find ferromagnetic arrangement among the impurity spins to be energetically favored over both paramagnetic and spin-density-wave ones. Such analysis based on Landau free energy is valid only in the low doping regime, and also applies for magnetic ordering on the surface of cubic topological Kondo insulators, when the effective chemical potential, defined in Eq.~(\ref{effective_chem}), is small. However, such analysis cannot not be extended to high doping regime and we have to rely on numerical analysis to gain insight into the magnetic ordering over a wide range of chemical doping.

Our central achievements from numerical analysis are displayed in Figs.~\ref{numerics_TI} and \ref{numerics_TKI}, respectively for simple topological insulators and cubic topological Kondo insulators. Irrespective of the doping level, the system always breaks into multiple small islands, each of which is ferromagnetically ordered. The size of such ferromagnetic grains $\ell \sim \mu^{-1}$ on the surface of topological insulator and $\ell \sim \mu^{-1}_{eff}$ on the surface of topological Kondo insulator. When the chemical doping is low the magnetization points in the same direction in these islands (but of different magnitude) and the system possesses net finite magnetization. Such ground state is referred to as a ferromagnet. By contrast, for high doping the direction of magnetization in those islands are randomly distributed and system possesses net zero magnetization. The ground state takes the form of glass, and we refer this phase as ferromagnetic spin glass. Similar conclusion also holds for the surface of cubic topological Kondo insulator, for which the effective chemical potential, defined in Eq.~(\ref{effective_chem}), plays the role of chemical potential. The spatial variation of magnetic moment on the surface of topological insulators can, for example, be detected by \emph{spin resolved} scanning tunneling microscope (STM).

By numerically computing the net magnetization one can also track the transition between pure and glassy ferromagnetic phases. As shown in Figs.~\ref{magnetization_TI} and ~\ref{magnetization_TKI}, when the chemical doping for the surface state is gradually increased there is a first order phase transition between these two phases around a critical chemical doping, for which the characteristic length scale for RKKY oscillation is $\sim$ average inter-impurity distance. Therefore, our proposed phases and the first order phase transition between distinct phases can be found on the surface of magnetically doped topological insulators by systematically tuning the surface chemical potential, which, for example, can be achieved by \emph{ionic liquid gating}~\cite{PaglioneLiquidgating} or by injecting non-magnetic ions.

Our analysis can also be consequential for the measurement of anomalous Hall effect on the surface of topological insulators. Recently it has been demonstrated through self-consistent calcualtion that when magnetic adatoms are arranged in ferromagnetic pattern, in turn they produces a \emph{mass} or gap for surface Dirac fermion, by \emph{globally} breaking the time-reversal-symmetry~\cite{EfimkinRKKY}. Such two component massive Dirac fermion naturally gives rise to anomalous Hall conductivity~\cite{haldane}, which, however is not quantized unless the chemical potential resides within the mass gap. Although in the high doping regime the system breaks into multiple islands and each such configuration produces massive Dirac fermion. In the low doping regime when magnetic moment in each such islands point in the same direction, the surface Dirac fermion can still remain massive. Presence of such ferromagnetism can lead to \emph{hysterysis} that has recently been observed in SmB$_6$~\cite{Paglioneedge}. Even inside the glassy phase, when magnetic moment of ferromagnetic island is randomly oriented, Dirac fermion acquires a spatially modulated mass. In particular, when two neighboring islands possess magnetic moments of opposite sign, the Dirac mass assumes the profile of a \emph{domain wall}, which supports \emph{one dimensional chiral edge state}~\cite{Jackiw, Semenoff}. Such chiral edge state can ultimately constitute a \emph{network} which may also give rise to finite anomalous Hall conductivity that has recently been observed on the surface of Bi$_2$Se$_3$~\cite{QAHERecentObservation}, a detailed analysis of which, however, goes beyond the scope of the present work, and remains as an interesting and challenging open problem (for discussion on similar issue see Ref.~\cite{Ohtsuki}).      

\begin{acknowledgements}
We thank  D. Efimkin, X. Li and X.P. Li for helpful discussions. This work is supported by JQI-NSF-PFC. B. R. is thankful to Nordita, Center for Quantum Materials for hospitality, where part of the manuscript was finalized.
\end{acknowledgements}

\appendix

\section{Static spin susceptibility for massless Dirac fremion}~\label{ren_app}

\begin{figure}
\begin{center}
\includegraphics[scale=0.33]{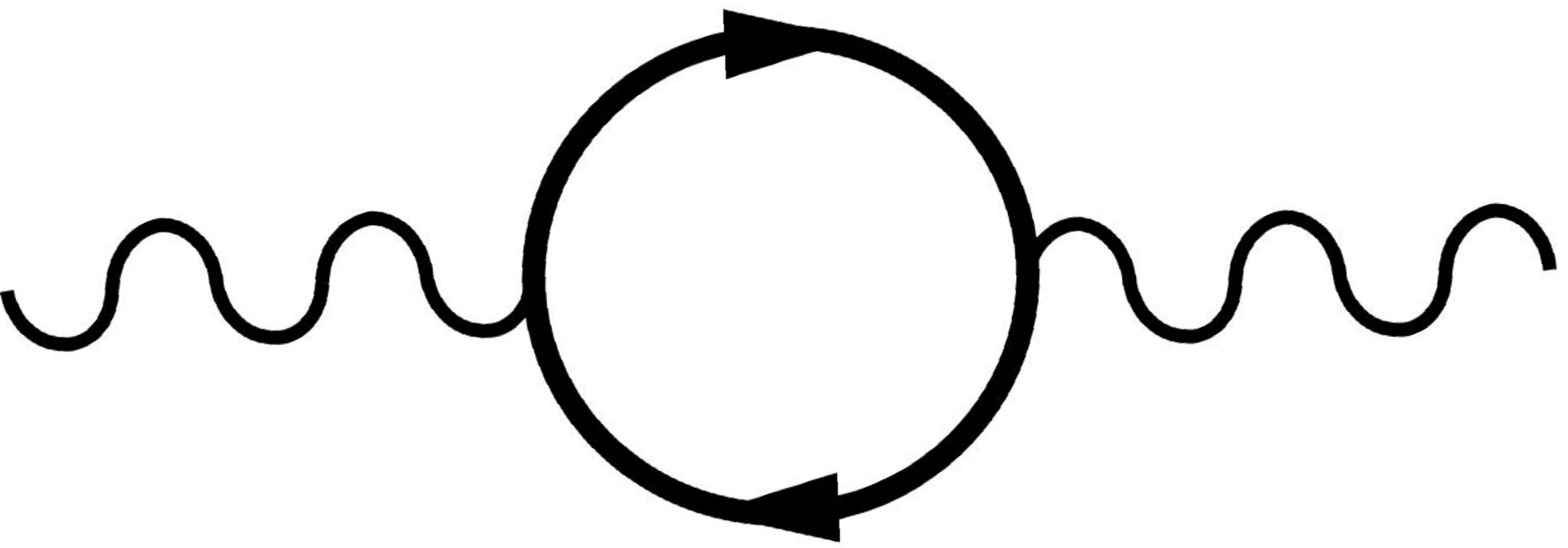}
\caption{Feynman diagram for spin susceptibility, where solid lines represent Dirac fermions and vertex is accompanied by Pauli matrix and wavy lines are external spin field. } 
\label{polarization}
\end{center}
\end{figure}

In this appendix, we derive the analytical expression for the static spin susceptibility after proper ultraviolet regularization, defined as $\chi^{ab}_{ren}(\bold{q}) = \chi^{ab}(\bold{q}) - \chi^{ab}(\bold{0})$. Its Feynman diagram is shown in Fig.~\ref{polarization}. From Eq.(\ref{sus_q}) we obtain
\begin{align}
&\chi^{ab}(\bold{q},iq_n) \nn
&=  \frac{1}{\beta} \sum_{i k_{n}} \int^{\prime} \frac{d^2\bold{k}}{(2\pi)^2} \mathrm{Tr} [\sigma^a \mathcal{G}(\bold{k} + \bold{q}, ik_n+iq_n ) \sigma^b \mathcal{G}(\bold{k} , ik_n ) ]. \nn \label{suscep-append}
\end{align}

The Green's function ($\mathcal{G}$) has already been defined in Eq.~(\ref{greens}). In terms of Feynman parameter ($x$) we can rewrite
\begin{align}
\frac{1}{[(ik_n+\mu)^2 - (\bold{k}+\bold{q})^2][(ik_n+\mu)^2 - \bold{k}^2]}  \nn 
= \int^1_0 dx \frac{1}{(\bold{k} + x\bold{q})^2 + (k_n-i\mu)^2 + \Delta^2]^2} 
\end{align}
where  $\Delta^2 = x(1-x)q^2$~\cite{Peskin}. Upon shifting the integral variable $\bold{k} + x\bold{q} \to \bold{k}$ the static spin susceptibility at zero temperature becomes
\begin{eqnarray}
\chi^{ab}( \bold{q} ) &=& \int^1_0 dx \int \frac{dk_0}{2\pi}  \int \frac{d^2\bold{k}}{(2\pi)^2} \frac{1}{[\bold{k} ^2 + (k_0 - i\mu)^2 + \Delta^2]^2} \nn
&\times& \mathrm{Tr}\bigg[ \sigma^a \big( ik_0 + \mu + [ \bold{k} +(1-x)\bold{q}] \times \bs{\sigma} \big) \nn 
&\times& \sigma^b \big( ik_0 + \mu + (\bold{k} -x\bold{q})\times \bs{\sigma} \big) \bigg] 
\label{sus_q_trace}
\end{eqnarray}

Let us first set $a=b=z$ in Eq.~(\ref{sus_q_trace}). We then obtain 
\begin{align}
\chi^{zz}(\bold{q},0) = -2\int^1_0dx\int \frac{dk_0}{2\pi} \int \frac{d\bold{k}}{(2\pi)^2} \frac{k^2_E - \Delta^2}{(k^2_E + \Delta^2)^2} ,
\end{align}
where  $k^2_E = \bold{k}^2 + (k_0 - i \mu)^2$. Notice that for $k_E \gg \Delta$, 
\begin{align}
\chi^{zz} \simeq \int^{\Lambda_{D}} d^3 k_E \frac{1}{k_E^2} \simeq \Lambda_{D}, 
\end{align}
where $\Lambda_{D}$ is the ultraviolet cutoff, and $\chi^{zz}$ display linear-$\Lambda_{D}$ divergence term. Such linear ultraviolet divergence in two spatial dimensions is a generic feature for low dimensional Dirac systems, dependence on which must be removed from any physical observable. By subtracting the $\bold{q}=0$ piece of $\chi^{zz}$, we finally arrive at the following renormalized quantity
\begin{align}
 \chi^{zz}_{ren}(\bold{q})  \equiv \chi^{zz}(\bold{q}) -  \chi^{zz}(\bold{0})
\end{align}
that is devoid of any $\Lambda_D$-dependence (for a different type of regularization in two dimensional relativistic systems, see Ref.~\cite{kennett}) and given by
\begin{align}
&\chi^{zz}_{ren}(\bold{q}) = \chi^{zz}(\bold{q}) -  \chi^{zz}(0) \nn 
& = -2\int^1_0dx\int \frac{dk_0}{2\pi} \int \frac{d\bold{k}}{(2\pi)^2} \Bigg[\frac{k^2_E - \Delta^2}{(k^2_E + \Delta^2)^2} - \frac{1}{k^2_E} \Bigg] \nn 
&= 2\int^1_0dx \Delta^2 \int \frac{dk_0}{2\pi} \int \frac{d\bold{k}}{(2\pi)^2} \frac{3k^2_E + \Delta^2}{(k^2_E + \Delta^2)^2k^2_E} \nn 
& =   \int^1_0 \frac{dx}{2\pi} \int \frac{dk_0}{2\pi} \left[ \frac{2\Delta^2}{ (k_0 - i \mu)^2+\Delta^2} + \log(1+\frac{\Delta^2}{ (k_0 - i \mu)^2}) \right] \nn 
& = \int^1_0 \frac{dx}{2\pi} \Theta(\Delta - \mu) \big[  \Delta + (\Delta- \mu) \big] \nn
& =  \int^1_0 \frac{dx}{2\pi} \Theta(\Delta - \mu) \big( 2 \Delta - \mu \big) ,\label{integral}
\end{align}
where $\Theta(x)$ is the step function. If the maximum of $\Delta  $ is less than $\mu$, i.e.
\begin{align}
\Delta_{max} = \sqrt{x_0(1-x_0)}q = \frac{q}{2} < k_F ,
\end{align}
then $ \chi^{zz}_{ren}(q < 2k_F) = 0$. On the other hand, for $q>2k_F$, we find
\begin{align}
\chi^{zz}_{ren}(\bold{q}) = \frac{q}{4\pi} \sin^{-1} \left[ 1-\frac{4k^2_F}{q^2} \right]^{1/2}.
\end{align}
where $x_{1,2} = \frac{1}{2} \mp \frac{1}{2}\sqrt{1-\frac{4k^2_F}{q^2}}$. When we combine the piecewise results together, we obtain the $zz$-component of the renormalized static spin susceptibility
\begin{align}
\chi^{zz}_{ren}(\bold{q}) = \frac{q}{4\pi} \mathrm{Re} \sin^{-1} \left[ 1-\frac{4k^2_F}{q^2}\right]^{1/2}.
\end{align}

On the other hand, for $a =b = x$ in Eq.(\ref{sus_q_trace}) we find
\begin{align}
& \chi^{xx}( \bold{q} ) \nn
& = -2 \int^1_0 dx \int \frac{dk_0}{2\pi}  \int \frac{d^2\bold{k}}{(2\pi)^2} \frac{ (k_0 - i \mu)^2 - \Delta^2 \cos 2\phi }{[\bold{k} ^2 + (k_0 - i\mu)^2 + \Delta^2]^2}  \nn 
& = - \int^1_0 \frac{dx}{2 \pi} \int \frac{dk_0}{2\pi}  \frac{ (k_0 - i \mu)^2 - \Delta^2 \cos 2\phi }{ (k_0 - i\mu)^2 + \Delta^2},
\end{align}
which also displays linear-$\Lambda$ divergence. Hence, we define the renormalized spin susceptibility as
\begin{align}
& \chi^{xx}_{ren}(\bold{q})=\chi^{xx}(\bold{q}) - \chi^{xx}(\bold{0})  \nn 
& = - \int^1_0 \frac{dx}{2 \pi} \int \frac{dk_0}{2\pi}  \frac{ -\Delta^2 - \Delta^2 \cos 2\phi }{ (k_0 - i\mu)^2 + \Delta^2}  \nn 
&=  \int^1_0 \frac{dx}{2 \pi} \Theta(\Delta - \mu) \Delta \cos^2 \phi = f_1 \cos^2 \phi,
\end{align}
where 
\begin{align}
f_1 = \frac{|k_F|}{4\pi} \mathrm{Re} \sqrt{1-\frac{4k^2_F}{q^2}} + \frac{q}{8\pi} \mathrm{Re} \sin^{-1} \sqrt{1-\frac{4k^2_F}{q^2}}.
\end{align}
Due to in plane rotational symmetry we find $\chi^{yy}_{ren}( \bold{q} ) = f_1 \sin^2 \phi$.

Next we compute the off diagonal elements of $\chi^{ab}$. For $a =x$ and  $b =y$ in Eq.(\ref{sus_q_trace}), we find 
\begin{eqnarray}
\chi^{xy}(\bold{q}) &=&  \int^1_0 dx \int \frac{dk_0}{2\pi}  \int \frac{d^2\bold{k}}{(2\pi)^2}
\frac{2 \Delta^2 \sin 2\phi }{[\bold{k} ^2 + (k_0 - i\mu)^2 + \Delta^2]^2}  \nn 
&=&   \int^1_0 \frac{dx}{2\pi} \int \frac{dk_0}{2\pi}   \frac{ \Delta^2 \sin 2\phi }{ (k_0 - i\mu)^2 + \Delta^2}  \nn 
& =& \frac{1}{2}\int^1_0 \frac{dx}{2\pi} \Theta(\Delta - \mu) \Delta \sin 2 \phi  = \frac{f_1}{2} \sin 2 \phi.
\end{eqnarray}
It is worth pointing out that $\chi^{xy}(\bold{q})$ is a ultraviolet finite quantity and $\chi^{yx}(\bold{q})=\chi^{xy}(\bold{q})$.  

Upon setting $a = x$ and $b = z$ in Eq.(\ref{sus_q_trace}) we obtain
\begin{align}
\chi^{xz}(\bold{q}) &= -\chi^{zx}(\bold{q}) =  \int^1_0 dx \int \frac{dk_0}{2\pi}  \int \frac{d^2\bold{k}}{(2\pi)^2}  \nn 
& \times \frac{ -2 (k_0 - i\mu) q \cos \phi }{[\bold{k} ^2 + (k_0 - i\mu)^2 + \Delta^2]^2} \nn
&  =  - \int^1_0 \frac{dx}{2\pi} \int \frac{dk_0}{2\pi}  \frac{ (k_0 - i\mu) q \cos \phi }{ (k_0 - i\mu)^2 + \Delta^2}.
\end{align}
Here, we need to construct a rectangular loop in the complex $k_0$-plane, one long side of the rectangle is $-\infty \to \infty$, the other $-\infty + i \mu \to \infty + i\mu $. Depending on whether the rectangle encloses the singular point $k_0 = i\mu - i \Delta$, we find
\begin{equation}
\int \frac{dk_0}{2\pi}  \frac{ (k_0 - i\mu) }{ (k_0 - i\mu)^2 + \Delta^2} = 0, \qquad \text{when} \quad \Delta > \mu,
\end{equation}
and 
\begin{align}
\int \frac{dk_0}{2\pi}  \frac{ (k_0 - i\mu) }{ (k_0 - i\mu)^2 + \Delta^2} = \frac{i}{2}, \qquad \text{when} \quad \Delta < \mu.
\end{align}
Then we obtain
\begin{align}
\chi^{xz}(\bold{q}) = - \chi^{zx}(\bold{q}) = -i f_2 \cos \phi
\end{align}
where
\begin{align}
f_2 = \frac{q}{4\pi} \left( 1 - \mathrm{Re} \sqrt{1-\frac{4k^2_F}{q^2}} \right)
\end{align}
Similarly, $\chi^{yz}(\bold{q}) = - \chi^{zy}(\bold{q}) = -i f_2 \sin \phi$. These off-diagonal entries do not depend on the ultraviolet cutoff.

\bibliography{Bib-RKKY}

\end{document}